\documentclass[a4paper,10pt]{article}

\usepackage[a4paper, total={6in, 10in}]{geometry}
\usepackage{amsmath,amssymb,amsfonts,amsthm}
\usepackage{bibentry}
\usepackage{mathrsfs,mathtools}
\usepackage{xcolor}
\usepackage{xspace}
\usepackage{hyperref}
\usepackage{euscript}
\usepackage[normalem]{ulem}
\usepackage[inline]{enumitem}
\usepackage{blindtext}
\usepackage[nottoc]{tocbibind}
\usepackage{caption}
\usepackage{subcaption}
\usepackage{picins}
\usepackage{authblk}
\usepackage{array}
\usepackage{url}
\usepackage{arydshln}
\usepackage{float}
\usepackage{csquotes}
\usepackage{datetime}

\DeclareRobustCommand{\legendline}[1]{\hspace{-0pt}\tikz[#1,line width=0.4mm,baseline=-0.5ex]{\draw (0,0) -- (.35,0);}\hspace{-0pt}}
\DeclareRobustCommand{\legendlined}[1]{\hspace{-0pt}\tikz[#1,line width=0.5mm,baseline=-0.5ex]{\draw[dotted] (0,0) -- (.3,0);}\hspace{-0pt}}

\definecolor{mblue}{rgb}{0,0.4470,0.7410}
\definecolor{morange}{rgb}{0.8500,0.3250,0.0980}
\definecolor{myellow}{rgb}{0.9290,0.6940,0.1250}
\definecolor{mpurple}{rgb}{0.4940,0.1840,0.5560}
\definecolor{mgreen}{rgb}{0.4660,0.6740,0.1880}
\definecolor{mcyan}{rgb}{0.3010,0.7450,0.9330}
\definecolor{mred}{rgb}{0.6350,0.0780,0.1840}
\definecolor{mgreenblue}{rgb}{0.0,1.0,0.5}
\definecolor{parulablue}{rgb}{0.2431,0.1490,0.6588}
\definecolor{parulalblue}{RGB}{39,151,235}
\definecolor{parulagreen}{RGB}{129,204,89}
\definecolor{parulayellow}{RGB}{249,251,21}
\definecolor{cblue}{rgb}{0,0.9,1}
\definecolor{corange}{rgb}{1,0.7,0}
\definecolor{mgray}{rgb}{0.8,0.8,0.8}

\newtheorem{applemma}{Lemma}[section]

\newcommand{\tss}[1]{\textsuperscript{#1}}
\newcommand{\lpvcore}{\textsc{LPVcore}\xspace}
\newcommand{\matlab}{\textsc{Matlab}\xspace}

\newcounter{ass}

\newcommand{\mc}[1]{\mathcal{#1}}
\newcommand{\mf}[1]{\mathfrak{#1}}
\newcommand{\mr}[1]{\mathrm{#1}}
\newcommand{\mb}[1]{\mathbb{#1}}

\newcommand{\mt}[1]{\mathtt{#1}}

\newcommand{\R}{\mathbb{R}}
\newcommand{\dnx}{n_\mr{x}}

\newcommand{\dnu}{n_\mr{u}}
\newcommand{\dnp}{n_\mr{p}}

\newcommand{\dnw}{n_\mr{w}}

\newcommand{\Xf}{X_+}

\newcommand{\Up}{U^\mt{p}}
\newcommand{\Xp}{X^\mt{p}}

\newcommand{\posdef}{\succ}
\newcommand{\negdef}{\prec}

\newcommand{\negsemidef}{\preccurlyeq}
\newcommand{\unaryminus}{\scalebox{0.65}[1]{\ensuremath{\,-}}}

\newcommand{\rankdef}[1]{\ensuremath{\mr{rank}\!\left(#1\right)}}

\newcommand{\diag}{\mr{diag}}
\newcommand{\rank}{\mr{rank}}

\newcommand{\kron}{\otimes} %

\newcommand{\gaminf}{\gamma}%
\newcommand{\gamtwo}{\gamma}%
\newcommand{\ltwo}{\ell_2}
\newcommand{\htwo}{\mc{H}^\mr{g}_2}
\newcommand{\Nd}{{N_\mr{d}}}
\newcommand{\dataset}{\mc{D}_\Nd}

\newcommand{\mcG}{\mathcal{G}}   %
\newcommand{\msc}[1]{\textsc{#1}}
\newcommand{\mcV}{\mc{V}}
\newcommand{\Z}{\mathbb{Z}}
\newcommand{\VP}{V}
\newcommand{\mcVpart}{V}
\newcommand{\mcVpartLTI}{\bar{\mcVpart}}  %
\newcommand{\mcVpartLPV}{\bar{\bar{\mcVpart}}}  %
\newcommand{\lyap}{P}
\newcommand{\onem}{1}
\newcommand{\Em}{E}
\newcommand{\Ef}{E_+}
\newcommand{\Zm}{Z}
\newcommand{\Zf}{Z_+}
\newcommand{\mcZf}{\mc{Z}_+}

\newlength\figH 
\newlength\figW
\usepackage{pgfplots}

\newtheorem{remark}{Remark}
\newtheorem{lemma}{Lemma}

\newtheorem{theorem}{Theorem}
\newtheorem{assumption}{Assumption}

\newtheorem{proposition}{Proposition}

\newtheorem{condition}{Condition}

\title{Direct Data-Driven State-Feedback Control of Linear Parameter-Varying Systems%
\thanks{Corresponding author Chris Verhoek. \texttt{c.verhoek@tue.nl}}%
}

\author[1]{Chris Verhoek}

\author[1,2]{Roland T{\'o}th}

\author[3]{Hossam S. Abbas}

\affil[1]{Control Systems group, Dept. of Electrical Engineering, Eindhoven University of Technology, Eindhoven, The Netherlands.}
\affil[2]{Systems and Control Lab, Institute for Computer Science and Control, Budapest, Hungary.}
\affil[3]{Institute for Electrical Engineering in Medicine, Universit{\"a}t zu L{\"u}beck, L{\"u}beck, Germany.}

\begin{document}

\maketitle

\begin{abstract}
    The framework of \emph{linear parameter-varying} (LPV) systems has shown to be a powerful tool for the design of controllers for complex nonlinear systems using linear tools. In this work, we derive novel methods that allow to synthesize LPV state-feedback controllers \emph{directly} from only a single sequence of data and guarantee stability and performance of the closed-loop system. %
We show that if the measured open-loop data from the system satisfies a persistency of excitation condition, then the full open-loop and closed-loop input-scheduling-state behavior can be represented using only the data. With this representation we formulate data-driven analysis and synthesis problems, where the latter yields controllers that guarantee stability and performance in terms of infinite horizon quadratic cost, generalized $\mc{H}_2$-norm and $\ltwo$-gain of the closed-loop system. The controllers are synthesized by solving a semi-definite program. Additionally, we provide a synthesis method to handle noisy measurement data. Competitive performance of the proposed data-driven synthesis methods is demonstrated w.r.t. model-based synthesis in multiple simulation studies, including a nonlinear unbalanced disc system.

\end{abstract}

\section{Introduction} \label{s:intro}
Due to increasing performance, environmental, etc., requirements,
 control design for the new generation of engineering %
systems is increasingly more challenging, %
as the dynamic behaviors of these systems are becoming %
dominated by nonlinear %
effects. A particularly interesting %
framework to deal with such challenges is the class of \emph{linear parameter-varying} (LPV) systems \cite{Toth2010}. %
LPV systems have a linear input-(state)-output relationship, but this relationship is dependent on a measurable, time-varying signal, referred to as the \emph{scheduling signal}. %
This scheduling signal is used to express the nonlinear/time-varying/exogenous components that are affecting the system, allowing to describe a wide range of nonlinear systems in terms of LPV surrogate models \cite{Toth2010}. In practice, LPV modeling and model-based control methods have been successfully deployed in many engineering problems, proving versatility of the framework to meet with the increasing complexity challenges and performance expectations~\cite{mohammadpour2011lpv}.

However, despite of the powerful LPV model-based control solutions, their deployment is becoming more challenging, as
modeling of the next generation of engineering systems via first-principles  is increasingly
 complex, time-consuming and often lacks sufficient accuracy. Hence,  engineers frequently need to obtain models based on data to analyze and control such systems. Despite the tremendous progress in LPV~\cite{Toth2010, Cox2021, van2008subspace} and nonlinear~\cite{ScLj19, BaoJavad2022} system identification, 
several  aspects of the system %
identification 
toolchain, such as
model structure selection, are either under-developed or are rather complex and demand several iterations. %
Moreover, research on \emph{identification for control} in the \emph{linear time-invariant} (LTI) case has shown that, to synthesize a controller to achieve a given performance objective in model-based control, only some dynamical aspects of the system are important~\cite{GEVERS2005335}. This means that focusing the identification process on obtaining only such information accurately for synthesis can achieve higher model-based control performance for a given experimentation budget. %
This lead to the idea that fusing the control objective into system identification and even accomplishing synthesis of a controller \emph{directly} from data, i.e., \emph{direct data-driven control}, carries many benefits \cite{skelton1994data, HouWang2013}, while it also simplifies the overall modeling and control toolchain {and thus opens up possibilities to automate the full control design process}. Based on this motivation, data-driven LPV control methods %
\cite{formentin2016direct, MillerSznaier2022, yahagi2022direct} have been introduced, which have provided promising solutions to the direct control design problem, however, often without guarantees on the stability and performance of the resulting closed-loop system.
 
In the LTI case, a cornerstone result, the so-called \emph{Fundamental Lemma} \cite{WillemsRapisardaMarkovskyMoor2005},  allows the design of \emph{direct} data-driven controllers with closed-loop stability and performance guarantees. Using this paradigm, numerous powerful results have been developed for LTI systems on data-driven simulation \cite{MarkovskyRapisarda2008, alsalti2024data}, stability and performance analysis \cite{romer2019one, dePersisTesi2020, vanWaarde2022_matrixS}, state-feedback control~{\cite{berberich2020robust, berberich2022combining}}  and predictive control~\cite{coulson2019data, BerberichKohlerMullerAllgower2021DPC_guarantees, verheijen2023handbook, breschi2023data}, many of which also provide robustness against noise or uncertainties. The extension of the Fundamental Lemma for the class of LPV systems has been derived in  %
\cite{VerhoekTothHaesaertKoch2021, VerhoekAbbasTothHaesaert2021} and it has already been used for the development of direct LPV data-driven predictive control design \cite{VerhoekAbbasTothHaesaert2021} and direct data-driven dissipativity analysis of LPV systems \cite{Verhoek2023_dissipativity} {using input-scheduling-output data. However, the data-driven LPV representation in the case of state measurements, as well as methods to achieve data-driven LPV state-feedback control synthesis using such a representation, has been missing from the literature so far. In this work, we derive novel, data-based methods for LPV state-feedback controller synthesis} that can provide closed-loop stability and performance guarantees based on just a single (short) sequence of measurement data from the system.
More specifically, our contributions in this paper are:

\begin{enumerate}[label={C\arabic*:}, align=left, ref={C\arabic*}, leftmargin=*]
	\item We derive an data-driven closed-loop LPV representation for a given controller from a single sequence of open-loop input-scheduling-state measurement data of an unknown LPV system that has a \emph{state-space} (SS) representation with affine scheduling dependence; \label{C:rep}
	\item We develop fully data-based analysis and synthesis methods of LPV state-feedback controllers to guarantee closed-loop stability. {Additionally, we extend these to include optimal performance in terms of quadratic, $\htwo$-norm and $\ltwo$-gain of the closed-loop system;} \label{C:stabperf}
	\item We extend the methods from Contribution~\ref{C:stabperf} to work under noisy measurements;\label{C:noise}
	\item We present an extensive set of simulation studies to demonstrate the capabilities of our methods, and show their advantages by comparing them to model-based LPV and data-driven LTI design methods. \label{C:sim}
\end{enumerate}

The direct data-driven analysis and synthesis techniques we present in this paper are formulated as a finite set of \emph{linear matrix inequalities} (LMIs), and thus can be efficiently solved as a \emph{semi-definite program} (SDP).

The paper is structured as follows. The problem setting is introduced in Section~\ref{s:problemstatement}, followed by the development of open- and closed-loop data-driven LPV representations in Section~\ref{s:datadrivenLPVrepresentation}, providing~\ref{C:rep}. Based on the derived representations, the direct LPV data-driven analysis and controller synthesis methods are formulated in Sections~\ref{s:datadrivenstabKLPV} and~\ref{s:datadrivenperf}, giving~\ref{C:stabperf}. We extend the results of Section~\ref{s:datadrivenstabKLPV} for the case of noisy data in Section~\ref{s:noise}, providing~\ref{C:noise}. We finish the paper with comparative simulation studies in Section~\ref{s:example}, demonstrating competitive performance of the proposed methods as part of~\ref{C:sim}. Finally, the conclusions are drawn in Section~\ref{s:conclusion}. To support the clarity and readability of our results, we have added a review of the existing model-based state-feedback analysis and synthesis methods in the appendix.

\subsubsection*{Notation}%
$\R$ denotes the set of real numbers, while the set of all integers is denoted by $\Z$. The set of real, symmetric $n\times n$ matrices is denoted as $\mb{S}^n$. For sets $\mb{A}$ and $\mb{B}$, $\mb{B}^\mb{A}$ indicates the collection of all maps from $\mb{A}$ to $\mb{B}$. The $n\times m$ zero matrix is denoted by $0_{n\times m}$, and we write $0_n$ when $m=n$. The identity matrix is denoted by $I_n$, while $\onem_{n}$ denotes the vector $\begin{bsmallmatrix} 1 & \cdots & 1 \end{bsmallmatrix}^\top\in\mb{R}^n$.  $X \succ 0$ and $X\prec 0$ ($X \succeq0 $ and $X \preceq 0$) denote positive/negative (semi) definiteness of a symmetric matrix $X\in\mb{S}^{n}$, respectively. The Kronecker product of $A\in\mathbb{R}^{n\times m}$ and $B\in\mathbb{R}^{p\times q}$ is denoted by $A\otimes B \in\mathbb{R}^{pm \times qn}$, and for $C\in\R^{m\times k}$, $D\in\R^{q\times l}$, the following identity holds~\cite{HoJo91}: 
\begin{equation}\label{kr:mult} (A\otimes B)(C\otimes D)=AC\otimes BD. \end{equation}
The Redheffer star product is $\Delta\star L$, which for $\Delta\in\mathbb{R}^{n\times m}$ and $L\in\mathbb{R}^{p\times q}$ with $n<p$ and $m<q$ gives the \emph{upper linear fractional
transformation} (LFT). Furthermore, the Moore-Penrose (right) pseudo-inverse is denoted by $\dagger$. We use $(*)$ to denote a symmetric term in a quadratic expression, e.g., $(*)^\top Q(a-b) = (a-b)^\top Q(a-b)$ for $Q\in\mathbb{R}^{n\times n}$ and $a,b\in\mathbb{R}^{n}$.  Block diagonal concatenation of matrices is given by $\mr{blkdiag}$, i.e., $\mr{blkdiag}(A,B) = \begin{bmatrix} A & 0 \\ 0 & B \end{bmatrix}$, $A\in\mathbb{R}^{n\times m}$, $B\in\mathbb{R}^{p\times q}$.

\section{Problem setting}\label{s:problemstatement}

Consider a discrete-time LPV system that can be represented by the following LPV-SS representation: 
\begin{subequations}\label{eq:LPVSS}
    \begin{align}
        x_{k+1} & = A(p_k)x_k+B(p_k)u_k, \label{eq:LPVSS:state}\\ 
        y_k & = x_k,
    \end{align}
\end{subequations}
where $k\in\mathbb{Z}$ is the discrete time, %
$x_k\in\R^{n_\mr{x}}$, $u_k\in\R^{n_\mr{u}}$, $y_k\in\R^{n_\mr{x}}$ and $p_k\in\mathbb{P}$ are  the state, input, output and scheduling signals, respectively, and $\mathbb{P}\subset\R^{\dnp }$ is a compact, convex set that defines the range of the scheduling signal.
The matrix functions $A:\mathbb{P}\rightarrow \mathbb{R}^{n_\mathrm{x}\times n_\mathrm{x}}$ and  $B:\mathbb{P}\rightarrow \mathbb{R}^{n_\mathrm{x}\times n_\mathrm{u}}$ are considered to have affine dependency on $p_k$, which is a common assumption in practice, cf. \cite{deLange2022lpv, Toth2011_SSrealizationTCST},
\begin{equation}\label{eq:LPVdependency}
    A(p_k)=A_0+\sum_{i=1}^{\dnp }p_{k,i}A_i, \qquad B(p_k)=B_0+\sum_{i=1}^{\dnp }p_{k,i}B_i,
\end{equation}
where $\{A_i\}_{i=1}^{\dnp }$ and $\{B_i\}_{i=1}^{\dnp }$ are real matrices with appropriate dimensions. The solutions of~\eqref{eq:LPVSS} are collected in the set 
\begin{equation}
    \mathfrak{B}=\{ (x,p,u)\in ( \R^{\dnx}\times \mathbb{P}\times  \R^{\dnu})^\mb{Z} \mid \text{\eqref{eq:LPVSS} holds } \forall k\in\mb{Z} \},
\end{equation}
which we refer to as the \emph{behavior}. To stabilize~\eqref{eq:LPVSS}, we can design an LPV state-feedback controller $K(p_k)$ corresponding to the control-law 
\begin{subequations}\label{eq:controllaw}
\begin{equation}\label{eq:controllaw:law}
    u_k = K(p_k)x_k,
\end{equation}
where we choose the LPV state-feedback controller $K:\mb{P}\to\mb{R}^{\dnu\times\dnx}$ to have affine dependence on $p_k$
\begin{equation}\label{eq:controllaw:dependencyK}
    K(p_k) = K_0 + {\textstyle\sum_{i=1}^{\dnp}}p_{k,i}K_i,
\end{equation}
\end{subequations}
similar to~\eqref{eq:LPVSS}. The well-known LPV state-feedback problem is to design $K(p_k)$ such that it ensures asymptotic (input-to-state) stability and minimizes a given performance measure (e.g.,~$\ltwo$-gain) of the closed-loop system
\begin{equation}\label{eq:closedloopsys}
    x_{k+1} = A_\msc{CL}(p_k)x_k= \left(A(p_k) + B(p_k)K(p_k)\right)x_k,
\end{equation}
under all scheduling trajectories $p\in\mb{P}^\mathbb{Z}$, where the solution set of~\eqref{eq:closedloopsys} is $\mf{B}_{\mr{CL}}=\mf{B}|_{u=K(p)x}\subseteq \pi_{\mr{x}, \mr{p}}\mf{B}$. 
For such a design of~$K(p_k)$, one will need the \emph{exact} mathematical description of~\eqref{eq:LPVSS}, which can be unrealistic in practical situations. In this paper, we consider the design of $K(p_k)$ for an unknown LPV system~\eqref{eq:LPVSS}, based \emph{only} on the measured data-set
\begin{equation}\label{eq:dataset}
    \dataset \coloneq \left\{u_k^\mr{d}, p_k^\mr{d}, x_k^\mr{d} \right\}_{k=1}^\Nd,
\end{equation}
which is a single trajectory from the open-loop LPV system~\eqref{eq:LPVSS}. This problem is formalized in the following problem statement.

\subsubsection*{Problem statement}
Consider a data-generating system that can be represented with~\eqref{eq:LPVSS}. Given the \emph{data-dictionary} $\dataset$ sampled from the data-generating system, how to synthesize an LPV state-feedback controller $K(p_k)$ based on only $\dataset$ such that it ensures stability and performance of the closed-loop~\eqref{eq:closedloopsys}?

In order to solve this problem, we first need to construct data-driven representations of $\mf{B}$ and $\mf{B}_{\mr{CL}}$, i.e., data-based realizations of~\eqref{eq:LPVSS} and~\eqref{eq:closedloopsys}, corresponding to Contribution~\ref{C:rep}.

\section{Data-Driven LPV Representations} \label{s:datadrivenLPVrepresentation}
The data-driven LPV-SS representations that are derived in this section are the key ingredients for the data-driven LPV state-feedback controller design methods that we present in Sections~\ref{s:datadrivenstabKLPV} and~\ref{s:datadrivenperf}.

\subsection{Open-loop data-driven LPV representation}
{In the literature, various forms of data-driven open-loop LPV-SS representations have been introduced already, see~\cite{dong2009closed, van2008subspace, gosea2018data, Cox2021}. Here we provide a brief overview of a representation form that we will generalize for the closed-loop controlled behavior that we will use to derive our analysis and synthesis results.}

First note that by separating the coefficient matrices in~\eqref{eq:LPVdependency} from the signals in~\eqref{eq:LPVSS}, we can rewrite the state equation~\eqref{eq:LPVSS:state} by introducing auxiliary signals $p_k\kron x_k$ and $p_k\kron u_k$:  %
\begin{equation}\label{eq:open-loop-model-based}
    x_{k+1} = \mc{A} \begin{bmatrix} x_k \\ p_k \kron x_k \end{bmatrix} + \mc{B}\begin{bmatrix} u_k \\ p_k\otimes u_k \end{bmatrix},
\end{equation}
where $\mc{A}=\begin{bmatrix} A_0 & A_1 & \cdots & A_{\dnp} \end{bmatrix}$ and $\mc{B}=\begin{bmatrix} B_0 & B_1 & \cdots & B_{\dnp} \end{bmatrix}$ are the aggregated coefficient matrices.
Then, given the measured data-dictionary $\dataset$ from~\eqref{eq:LPVSS}, the construction of 
the following matrices
\begin{subequations}\label{eq:data-matrices}
\begin{align}
	U & = \begin{bmatrix} u^\mr{d}_1 & \cdots  & u^\mr{d}_{\Nd-1} \end{bmatrix}\in\R^{n_\mr{u}\times \Nd-1}, & \Up & = \begin{bmatrix} p^\mr{d}_1\otimes u^\mr{d}_1 & \cdots &  p^\mr{d}_{\Nd-1}\otimes u^\mr{d}_{\Nd-1} \end{bmatrix}\in\R^{\dnp n_\mr{u}\times \Nd-1}, &&\\ 
	X & = \begin{bmatrix} x^\mr{d}_1     & \cdots  & x^\mr{d}_{\Nd-1} \end{bmatrix}\in\R^{n_\mr{x}\times \Nd-1}, & \Xp & = \begin{bmatrix} p^\mr{d}_1\otimes x^\mr{d}_1 & \cdots &  p^\mr{d}_{\Nd-1}\otimes x^\mr{d}_{\Nd-1} \end{bmatrix}\in\R^{\dnp n_\mr{x}\times \Nd-1}, && \\
	\Xf & = \begin{bmatrix} x^\mr{d}_2 &  \cdots  & x^\mr{d}_{\Nd}\end{bmatrix}\in\R^{n_\mr{x}\times \Nd-1}, &&&
\end{align}
\end{subequations}
allows to write the relationship between $\mc{A},\mc{B}$ and the data in $\dataset$ as
\begin{equation}\label{eq:LPVSysIdent}
	\Xf=\mc{A}
	\begin{bmatrix}
	X\\ \Xp
	\end{bmatrix}+
	\mc{B} \begin{bmatrix}
	U\\ \Up
	\end{bmatrix}.
\end{equation}
Note that~\eqref{eq:LPVSysIdent} holds true due to the linearity property of LPV systems \emph{along} a given $p$ trajectory. This is a well-known fact and often used in LPV subspace identification, cf.~\cite{Cox2021}. Instead of estimating $\mc{A},\mc{B}$ (as is done in LPV system identification), we now \emph{represent} the LPV system given by~\eqref{eq:LPVSS} with $\dataset$ using the data matrices in~\eqref{eq:data-matrices}:
\begin{equation}\label{eq:open-loop-data-based}
	x_{k+1} = \Xf \mcG^\dagger \begin{bmatrix} x_k \\ p_k\otimes x_k \\ u_k \\ p_k\otimes u_k
	\end{bmatrix}, \quad \mcG:= \begin{bmatrix} X \\ \Xp \\ U \\ \Up \end{bmatrix}.
\end{equation}
{We have now characterized the open-loop signal relationship of the state transition~\eqref{eq:open-loop-model-based} using \emph{only} data, giving an open-loop data-driven LPV representation. Before we extend this to a data-driven closed-loop representation, it is important to discuss its well-posedness.} 

The data-based representation~\eqref{eq:open-loop-data-based} is \emph{well-posed} under the condition that the data-set $\dataset$ is \emph{persistently exciting} (PE).
Note that the PE condition is always defined with respect to a certain system class/representation, which is in this case a static-affine LPV-SS representation with full state observation. Contrary to the PE condition for shifted-affine LPV-IO representations~\cite{VerhoekTothHaesaertKoch2021, newpaper}, the PE condition for the system class considered in this paper is simpler, and directly corresponds to the existence of the right pseudo-inverse $\mcG^\dagger$ of $\mcG$,
giving the following PE condition for $\dataset$ to imply existence of~\eqref{eq:open-loop-data-based}:

\begin{condition}[Persistency of Excitation]\label{cond:rank-cond}
	If $\mcG$ %
	has full \emph{row} rank, i.e., $\rankdef{\mcG}=(1+\dnp )(n_\mr{x}+n_\mr{u})$, then $\dataset$ is persistently exciting w.r.t.~\eqref{eq:LPVSS}~and~\eqref{eq:LPVdependency}.%
\end{condition}
Based on this condition, we need at least $\Nd \geq 1+(1+\dnp )(n_\mr{x}+n_\mr{u})$ time samples for the construction of a well-posed~\eqref{eq:open-loop-data-based}. Moreover, as we assume full state observation, the parameters $\dnx, \dnp, \dnu$ are also known and thus no additional conditions are required to ensure existence of~\eqref{eq:open-loop-data-based}. {Note that Condition~\ref{cond:rank-cond} is an \emph{a posteriori} check on $\dataset$. Ensuring this condition \emph{a priori} by experiment design largely depends on the exact nature of $p$ in relation with the represented physical system ($p$ is, e.g., a free, external, physical variable, or related to variables such as states/inputs/outputs in case of the LPV embedding of a nonlinear system) and hence it is not explored further here.

Additionally, Condition~\ref{cond:rank-cond} is also connected to the generalized PE condition of~\cite{newpaper}, which is a condition on the data that verifies whether the full length-$L$ \emph{input-output} behavior of an LPV system under shifted-affine scheduling dependency can be represented. Here, we have that $L=1$ and the output dimension is $\dnx$. For this case, the generalized PE condition of~\cite{newpaper} \emph{coincides} with Condition~\ref{cond:rank-cond}.}

\subsection{Closed-loop data-driven LPV representation} \label{sec:cl:ddrep}
{Even though the data-driven representation of the open-loop case has been investigated in the literature, the same has not been formulated for the closed-loop system~\eqref{eq:closedloopsys}. Especially not for the case when both the system representation~\eqref{eq:LPVSS} and the feedback controller~\eqref{eq:controllaw} have affine dependence on $p$. 
The following theorem gives conditions such that we can represent~\eqref{eq:closedloopsys} using \emph{only} the data-set $\dataset$ taken from the open-loop system~\eqref{eq:LPVSS}, i.e., this result provides a data-based parameterization of the closed-loop system for a given state-feedback law~\eqref{eq:controllaw}. Note that later we will use this representation for the synthesis problem, where the controller is \emph{not} known on beforehand.}
\begin{theorem}[Data-based closed-loop representation]\label{th:closed-loop-data-based-general}
	Given the data-dictionary $\dataset=\{u_k,p_k,x_k\}_{k=1}^{\Nd}$, measured from~\eqref{eq:LPVSS} that satisfies  Condition~\ref{cond:rank-cond}. %
	Let $\Xf$ and $\mcG$ be defined as in~\eqref{eq:data-matrices} and~\eqref{eq:open-loop-data-based}
	 under $\dataset$. For an LPV controller $K$ given by~\eqref{eq:controllaw},  the closed-loop system~\eqref{eq:closedloopsys} is represented equivalently as
	\begin{equation}\label{eq:data-based-CLLPV-state-feedback-general}
		x_{k+1}=\Xf \mcV \begin{bmatrix}
		x_k\\p_k\otimes x_k \\ p_k\otimes p_k\otimes x_k
		\end{bmatrix},
	\end{equation} 
	where $\mcV\in\mathbb{R}^{\Nd-1 \times n_\mathrm{x}(1+n_\mathrm{p}+n_\mathrm{p}^2) }$ is any matrix that satisfies
	\begin{equation}\label{eq:consist-cond-general}
		\mathcal{M}_\textsc{CL} = \mcG\mcV, \quad \text{where } \mathcal{M}_\textsc{CL}:=\begin{bmatrix}
		I_{n_\mr{x}} & 0& 0\\
		0& I_{\dnp }\otimes I_{n_\mr{x}}& 0\\
		K_0 & \bar{K} & 0\\
		0 & I_{\dnp }\otimes K_0 &  I_{\dnp }\otimes \bar{K} 
		\end{bmatrix}.
	\end{equation}
\end{theorem} 
\begin{proof}
    If Condition~\ref{cond:rank-cond} holds, then, by the Rouch{\'e}-Capelli Theorem, there always exists a $\mcV$ such that~\eqref{eq:consist-cond-general} holds. Now, let us write the controller~\eqref{eq:controllaw} as 
    \begin{equation}\label{eq:pfth1:1}
        u_k = \mc{K} \begin{bmatrix} x_k\\p_k\otimes x_k \end{bmatrix}, \quad
    		\mc{K} = \begin{bmatrix} K_0 & K_ 1 & \cdots & K_{\dnp } \end{bmatrix}.
    \end{equation}
    When~\eqref{eq:pfth1:1} is substituted into~\eqref{eq:closedloopsys}, we obtain the closed-loop
    \begin{equation}\label{eq:clsub}
        x_{k+1} = \begin{bmatrix} A_0 & \bar{A} \end{bmatrix} \begin{bmatrix} x_k \\ p_k \kron x_k \end{bmatrix} + B_0\mc{K} \begin{bmatrix} x_k \\ p_k \kron x_k \end{bmatrix} + \bar{B} \cdot \big(p_k \kron(K_0x_k + \bar{K}\cdot(p_k\kron x_k))\big),
    \end{equation}
    where $\bar{A} = \begin{bmatrix} A_1 & \cdots & A_{\dnp } \end{bmatrix}$, and $\bar{B},\bar{K}$ are similarly defined. %
    Using the Kronecker property~\eqref{kr:mult}, we can rewrite~\eqref{eq:clsub} as
    \begin{equation}\label{eq:pfth1:2}
        x_{k+1} = \underbrace{\begin{bmatrix} A_0 + B_0K_0 \ & \ \bar{A} + B_0\bar{K} + \bar{B}(I_{\dnp}\kron K_0) \ & \ \bar{B}(I_{\dnp}\kron\bar{K}) \end{bmatrix}}_{\mc{M}} \begin{bmatrix} x_k\\p_k\kron x_k \\ p_k\otimes p_k\otimes x_k 	\end{bmatrix},
    \end{equation}
    where $\mc{M}\in\mb{R}^{\dnx\times\dnx(1+\dnp+\dnp^2)}$ now fully defines the relationship between the signals in the closed-loop LPV system. Note that $\mc{M}$ can be rewritten as
	\begin{equation}\label{eq:proof1}
		\mc{M} = \begin{bmatrix} \mc{A} & \mc{B} \end{bmatrix} \mc{M}_\textsc{CL}.
	\end{equation}
	Substituting the relation $\begin{bmatrix} \mc{A} & \mc{B}\end{bmatrix} = \Xf\mcG^\dagger$ and~\eqref{eq:consist-cond-general}
	into~\eqref{eq:proof1} yields
	\begin{equation}\label{eq:pfth1:3}
	    \mc{M}= \Xf\mcV,
	\end{equation}
	which, through~\eqref{eq:clsub}, gives the data-based closed-loop representation~\eqref{eq:data-based-CLLPV-state-feedback-general}, equivalent to the model-based closed-loop representation~\eqref{eq:closedloopsys}. 
\end{proof}
{Note that, in terms of the data, Theorem~\ref{th:closed-loop-data-based-general} gives an \emph{implicit} data-driven parametrization of the closed-loop for a given controller $K(p_k)$ and measured open-loop data.} Under Condition~\ref{cond:rank-cond} and a {given} parametrization $\mc{K}$, we can also compute an \emph{explicit} data-driven LPV representation of the closed-loop via $\mcV= \mcG^\dagger \mathcal{M}_\textsc{CL}$. However, this would be \emph{only a particular} (minimum 2-norm solution) of~\eqref{eq:consist-cond-general}, while Theorem~\ref{th:closed-loop-data-based-general} allows for an affine subspace of solutions in terms of $\mcV$ in the orthogonal projection of $\mathcal{M}_\textsc{CL}$ on the range of $\mcG$. 

Furthermore, for a given $\mcV$, {but unknown $\mc{K}$},  
condition~\eqref{eq:consist-cond-general} allows %
to \emph{recover} a controller $K(p_k)$, which is the key component for controller synthesis. To this end, partition $\mcV$ as $\mcV:=\begin{bmatrix} \mcVpart_0 & \mcVpartLTI & \mcVpartLPV \end{bmatrix}$ {according to the signal dimensions in~\eqref{eq:pfth1:2}}, giving
\[
	\mcVpartLTI = \begin{bmatrix} \mcVpart_1 & \cdots & \mcVpart_{\dnp } \end{bmatrix}, \quad
	\mcVpartLPV = \begin{bmatrix} \mcVpart_{\dnp +1} & \cdots & \mcVpart_{n_\mr{p}^2} \end{bmatrix}, \qquad V_i \in\mb{R}^{\Nd-1\times \dnx}.
\]
Then, based on~\eqref{eq:consist-cond-general}, we can derive that
\begin{equation*}
	K_0=UV_0,\quad \bar{K}=U\mcVpartLTI.
\end{equation*}
Hence, based on a particular choice of $\mcV$ that satisfies~\eqref{eq:pfth1:3}, the corresponding control law can be recovered, and is fully defined in terms of the data $\dataset$ as
\begin{equation*}
	u_k = U \begin{bmatrix} \mcVpart_0 & \mcVpartLTI \end{bmatrix}
			\begin{bmatrix} 	x_k \\ p_k\otimes x_k \end{bmatrix}.
\end{equation*}
Note that, in order to represent the \emph{control law}, the term $\mcVpartLPV$ is redundant and only required to fulfill the data relations in~\eqref{eq:consist-cond-general}. We will now derive synthesis algorithms that allow us to \emph{find} realizations of $\mcV$ {(and thus $K$) that correspond to stable closed-loop representations}, and thus synthesize stabilizing LPV state-feedback controllers using only the information in $\dataset$.

\section{Data-driven synthesis of stabilizing LPV controllers}\label{s:datadrivenstabKLPV}
Based on the data-driven representation of the closed-loop system developed in Section \ref{sec:cl:ddrep}, we will derive data-driven controller synthesis methods %
that can be solved as an 
SDP. We show that these methods synthesize controllers that ensure closed-loop stability and achieve a wide range of performance targets (the latter will be discussed in Section~\ref{s:datadrivenperf}). We first provide an essential tool that we use to make the analysis and synthesis tractable.
\subsection{Full-block $\mc{S}$-procedure}
    We first give the full-block $\mc S$ procedure from \cite{Sc01,WuDo06}. This result is instrumental and used extensively throughout the paper. As it is often done in the literature, in the sequel, for conditions such as in the following lemma, we will use the notation `$p$' for both the scheduling \emph{signal} $p:\Z\rightarrow \mb{P}$ and for constant \emph{vectors} $p\in\mb{P}$ to describe all possible values $p_k$, i.e., the value of the signal $p$ at time moment $k\in\Z$ can take, similarly for $x$ and $u$. Where possible confusion might arise, we will clarify in the text which notion we refer to. 
\begin{lemma}[Full-block $\mc{S}$-procedure \cite{Sc01,WuDo06}]\label{lem:full-block-S-procedure}
	 Given a quadratic matrix inequality
	\begin{equation}\label{eq:quad-matrix-inequality}
		L^\top(p)W L(p)\posdef0, \quad \forall p \in\mathbb{P},
	\end{equation}
	where $L(p)=\Delta_{p}\star \bar{L} =L_{22}+L_{21}\Delta_{p}(I-L_{11}\Delta_{p})^{-1}L_{12}$, with
	\begin{equation} 
	   \Delta_{p}=\mr{blkdiag}\big(p_1 I_{n_{\Delta 1}},\ p_2 I_{n_{\Delta 2}},\cdots,p_{\dnp} I_{n_{\Delta \dnp}}\big),
	\end{equation}
	and $\mathbb{P}$ is convex. Then~\eqref{eq:quad-matrix-inequality} holds if and only if there exists a real full-block multiplier $\Xi\in\mb{S}^{2(\sum_{i=1}^{\dnp} n_{\Delta i})}$, defined as
    $\Xi=\begin{bsmallmatrix}\Xi_{11} & \Xi_{12} \\ \Xi_{12}^\top & \Xi_{22} \end{bsmallmatrix}$,
	 such that
	 \begin{subequations}\label{eq:quad-matrix-inequality-multipliers}
    \begin{align}
			\left[\begin{array}{c}
			* \\ \hline  *
			\end{array}\right]^\top	 
			\left[\begin{array}{c|c}
			\Xi & 0 \\ \hline  0 & - W
			\end{array}\right]	
			\left[\begin{array}{c c}
			L_{11} & L_{12} \\ I & 0 \\ \hline  L_{21} & L_{22} 
			\end{array}\right]\prec 0 , \label{eq:quad-matrix-inequality-multipliers:1}\\
			\left[\begin{array}{c}
			* \\ \hline  *
			\end{array}\right]^\top \Xi
			\left[\begin{array}{c}
			I \\ \hline  \Delta_{p}
			\end{array}\right]  \succeq 0, \label{eq:quad-matrix-inequality-multipliers:2}
    \end{align}
    for all $p\in\mb{P}$. The additional condition
    \begin{equation}
        \Xi_{22}\negdef 0
    \end{equation}
    yields~\eqref{eq:quad-matrix-inequality-multipliers:2} convex in $p$ (at the cost of necessity).
    \end{subequations}
\end{lemma}
Note that if $\mb P$ is a convex set that can be represented as the convex hull of finite many generators $\mt{p}^{i}$, i.e., it is a polytope given as $\mb{P} = \mr{co}(\{\mt{p}^{i}\}_{i=1}^{n_\mathrm{v}})$, where $\mr{co}$ denotes the convex hull, then verifying~\eqref{eq:quad-matrix-inequality-multipliers} for all $p\in\mb{P}$ reduces to verifying~\eqref{eq:quad-matrix-inequality-multipliers} only for the generators $\{\mt{p}^{i}\}_{i=1}^{n_\mathrm{v}}$, corresponding to an SDP.

\subsection{Closed-loop data-driven stability analysis} \label{ss:stabanalysis}
We first solve the direct data-driven stability analysis problem from which the synthesis methods for stabilizing LPV state-feedback control can be derived later. 
Quadratic asymptotic stability of the LPV system given by~\eqref{eq:LPVSS}, i.e., boundedness and convergence of the state-trajectories to the origin under $u\equiv 0$, is implied with the existence of a (positive definite and decrescent) Lyapunov function $V(x)=x^\top \lyap^{-1} x > 0$, $\forall x\in\R^{\dnx}\setminus\{0\}$, with $\lyap^{-1}\in\mb{S}^{\dnx}$ that satisfies $V(x_{k+1})-V(x_k)<0$ under all $(x,p,0)\in\mf{B}$. Working this out for a given state-feedback controller $K(p_k)$ as in~\eqref{eq:controllaw}, we obtain the well-known condition to analyze closed-loop stability in a model-based sense, see, e.g., Lemma~\ref{lem:Lyapunov-LPV-model-based} or~\cite{rotondo2015linear}. That is, the closed-loop LPV system~\eqref{eq:closedloopsys} is quadratically, asymptotically stable if 
\begin{equation}\label{eq:Lyapunov-LPV-model-based-2}
   \begin{bmatrix} \lyap & \lyap M^\top(p)  \\  M(p)\lyap & \lyap \end{bmatrix} \succ 0, \quad M(p) = \mc{M} \begin{bmatrix} I_{\dnx}\\ p\kron I_{\dnx} \\ p\kron p\kron I_{\dnx} \end{bmatrix},
\end{equation}
for all $p\in\mathbb{P}$ with\footnote{Choosing a parameter-varying Lyapunov function  can reduce conservatism of the analysis, but causes~\eqref{eq:Lyapunov-LPV-model-based-2} to have 3\tss{rd}-order polynomial dependency on $p$, making it difficult to arrive to an SDP form of the analysis and synthesis problems. See~\cite{Pereira2021} for a possible extension.} 
$\lyap\posdef0$. 
\begin{remark}[Checking stability over $\mb{P}$]
    {Satisfying~\eqref{eq:Lyapunov-LPV-model-based-2} for all $p\in\mb{P}$ can be recasted to an LMI that is convex in $p$ using, e.g., Lemma~\ref{lem:full-block-S-procedure}. Then, if $\mb P$ is a convex set generated by finitely many vertices, i.e., it is a polytope,  checking~\eqref{eq:Lyapunov-LPV-model-based-2} for all $p\in\mb P$ reduces to checking the feasibility of the LMI~\eqref{eq:Lyapunov-LPV-model-based-2} on the generators of $\mb{P}$ and hence can be efficiently solved as an SDP~\cite{SchererWeiland2021}. As an alternative, one can define a dense grid over $\mb{P}$ and solve~\eqref{eq:Lyapunov-LPV-model-based-2} for every point on the grid, which only guarantees quadratic asymptotic stability in a neighborhood of the grid points. In this paper, we only consider the former approach due to its global guarantees. However, the results in this paper can easily be formulated using a grid-based synthesis.}
\end{remark}
The formulation in~\eqref{eq:Lyapunov-LPV-model-based-2} allows us to use the closed-loop representation in Theorem~\ref{th:closed-loop-data-based-general} to derive the following theorem, which provides a computable method to analyze the stability of the unknown LPV system in closed-loop with $K(p_k)$ in a fully data-driven setting.

\begin{theorem}[Data-driven feedback stability analysis] \label{thm:stabilizing-sch-dependence}
	Given a data-set $\dataset$,  satisfying Condition~\ref{cond:rank-cond}, from a system that can be represented by~\eqref{eq:LPVSS}.
	For a $\lyap\in\mb{S}^{\dnx}$, let $\mc{F}\in\mb{R}^{\Nd-1\times\dnx(1+\dnp+\dnp^2)}$ and $F_Q\in\mb{R}^{(\Nd-1)(1+\dnp)\times \dnx (1+\dnp)}$ be defined such that
	\begin{equation}\label{eq:F=VZ}
	    F(p) := \mc{V} \begin{bmatrix} I_{\dnx}\\ p\kron I_{\dnx} \\ p\kron p\kron I_{\dnx} \end{bmatrix}\lyap = \mc{F} \begin{bmatrix} I_{\dnx}\\ p\kron I_{\dnx} \\ p\kron p\kron I_{\dnx} \end{bmatrix} = \begin{bmatrix} I_{\Nd-1} \\ p\otimes I_{\Nd-1} \end{bmatrix}^{\top} F_Q \begin{bmatrix} I_{n_\mr{x}} \\  p\otimes I_{n_\mr{x}} \end{bmatrix}.
	\end{equation}
	Then,
	the LPV state-feedback controller $K(p)$  stabilizes~\eqref{eq:LPVSS} under the feedback-law~\eqref{eq:controllaw}, if there exists an $F_Q$, a $\lyap\posdef0$ 
	and a multiplier $\Xi\in\mb{S}^{4\dnp\dnx}$,
	which satisfy
	\begin{equation}\label{eq:coupling-condition-sch-dep-4}
    	\mathcal{M}_\textsc{CL}
    	\begin{bmatrix}
    	{\lyap}&0&0\\0&I_{\dnp }\otimes {\lyap}&0\\0&0&I_{\dnp }\otimes I_{\dnp }\otimes {\lyap}
    	\end{bmatrix}=\mcG \mathcal{F},
    \end{equation}
	for the given data dictionary and also satisfy the LMI conditions in~\eqref{eq:quad-matrix-inequality-multipliers} for all $p\in\mb{P}$, with $\Delta(p) = \diag(p)\otimes I_{2n_\mr{x}}$, and where $W$ and $L_{11},\dots,L_{22}$ are given by: 
	\begin{align}\label{eq:LFT-matrices}
	   \begin{array}{rcl}
    	    W & = & \begin{bmatrix}
    	               \lyap_0 & \mc{\Xf}F_Q \\ 
    	               (\mc{\Xf}F_Q)^\top & \lyap_0
    	           \end{bmatrix}, \\ 
    	    \lyap_0 & = & \mr{blkdiag}(\lyap,0_{\dnx\dnp}),\vphantom{\Bigg)} \\ 
    	    {\mc{\Xf}} & = & \mr{blkdiag}(\Xf, I_{\dnp}\kron\Xf), 
	    \end{array} \quad \begin{array}{rclrcl}
		  L_{11} & = & 0_{2\dnx\dnp}, \vphantom{\Big)}& 
		  L_{12} & = & \onem_{\dnp}\otimes I_{2\dnx},\\
    	  L_{21} & = & \begin{bmatrix} 0_{\dnx\times 2\dnx\dnp} \\  I_{\dnp}\otimes \begin{bmatrix} I_{\dnx} & 0\end{bmatrix} \\ 
    	               0_{\dnx\times2\dnx\dnp} \\ 
    	               I_{\dnp}\otimes \begin{bmatrix} 0 & I_{\dnx} \end{bmatrix}
    		        \end{bmatrix}, & \
          L_{22} & = & \begin{bmatrix} \begin{bmatrix} I_{\dnx} & 0\end{bmatrix} \\ \onem_{\dnp}\otimes 0_{\dnx\times2\dnx} \\ \begin{bmatrix} 0 & I_{\dnx} \end{bmatrix} \\
    		           \onem_{\dnp}\otimes 0_{\dnx\times 2\dnx} \end{bmatrix}.
        \end{array}
    \end{align}
    If $\mb{P} = \mr{co}(\{\mt{p}^{i}\}_{i=1}^{n_\mathrm{v}})$ then~\eqref{eq:quad-matrix-inequality-multipliers} with \eqref{eq:LFT-matrices} is only required to be satisfied on $\{\mt{p}^{i}\}_{i=1}^{n_\mathrm{v}}$, corresponding to an SDP.
\end{theorem}
\begin{proof}
Substituting the relation~\eqref{eq:proof1} of the data-based closed-loop representation into~\eqref{eq:Lyapunov-LPV-model-based-2} results in
\begin{equation}\label{eq:ddLyap-1-general}
\begin{bmatrix}
\lyap & \lyap\VP^\top(p)\Xf^\top  \\  \Xf\VP(p)\lyap & \lyap
\end{bmatrix} \succ 0,
\quad \forall p\in\mathbb{P}, 
\end{equation}
where\footnote{Note that $\VP$ is unrelated to the Lyapunov function.}
\[ \VP(p) = \mcV \begin{bmatrix} I_{\dnx}\\ p\kron I_{\dnx} \\ p\kron p\kron I_{\dnx} \end{bmatrix}, \]
and $\mcV$ is restricted by~\eqref{eq:consist-cond-general}, corresponding to $\mathcal{M}_\textsc{CL}=\mcG\mcV$. To this end, let us introduce the matrix function $F(p):=\VP(p)\lyap$, resembling~\eqref{eq:F=VZ}. Substituting $F(p)$ in~\eqref{eq:ddLyap-1-general} results in
\begin{equation}\label{eq:ddLyap-1-general-3}
    \begin{bmatrix} \lyap & (\Xf F(p))^\top  \\  \Xf F(p) & \lyap \end{bmatrix} \succ 0, \quad \forall p\in\mathbb{P},
\end{equation}
while the substitution of $F(p)$ in~\eqref{eq:consist-cond-general} yields
\begin{equation}\label{eq:coupling-condition-sch-dep}
    \mathcal{M}_\textsc{CL}\begin{bsmallmatrix}
    I_{n_\mr{x}}\\p\otimes I_{n_\mr{x}}\\ p\otimes p\otimes I_{n_\mr{x}}
    \end{bsmallmatrix} \lyap =\mcG F(p),
\end{equation}
which \emph{couples} $\lyap$ with condition~\eqref{eq:consist-cond-general}. Note that~\eqref{eq:ddLyap-1-general-3} is not an LMI, due to the quadratic dependence of $F$ on $p$. Using the Kronecker property~\eqref{kr:mult},
    we can simplify~\eqref{eq:coupling-condition-sch-dep} as
    \begin{equation}\label{eq:pfth2:1}
    	\mathcal{M}_\textsc{CL}
    	\begin{bmatrix}
    	\lyap  &0&0\\0&I_{\dnp }\otimes \lyap &0\\0&0&I_{\dnp }\otimes I_{\dnp }\otimes \lyap  
    	\end{bmatrix} \begin{bmatrix}
    	I_{n_\mr{x}}\\p_k\otimes I_{n_\mr{x}}\\ p_k\otimes p_k\otimes I_{n_\mr{x}}
    	\end{bmatrix}
    	 =\mcG F(p_k).
    \end{equation}
    This immediately reveals the required scheduling dependency of $F(p_k)$ that is inherited from the left-hand side, cf.~\eqref{eq:F=VZ}, which allows us to further simplify~\eqref{eq:pfth2:1} to~\eqref{eq:coupling-condition-sch-dep-4}. Now, consider the definition of $F_Q$ in~\eqref{eq:F=VZ}. With the application of Kronecker property~\eqref{kr:mult} twice on $\Xf F(p)$, we can write it~as
\begin{equation}\label{eq:XarrowF(p)-general-quadratic}
	\Xf F(p)= \begin{bmatrix} 	I_{n_\mr{x}} \\  p\otimes I_{n_\mr{x}} \end{bmatrix}^{\top} \mc{\Xf} F_Q  \begin{bmatrix} I_{n_\mr{x}} \\  p\otimes I_{n_\mr{x}} \end{bmatrix},
\end{equation}
which is a quadratic form. This allows to write~\eqref{eq:ddLyap-1-general-3} as~\eqref{eq:quad-matrix-inequality} with $W$ as in~\eqref{eq:LFT-matrices} and
\(L(p) = \mr{blkdiag}\left( \begin{bsmallmatrix} I_{\dnx} \\ p\kron I_{\dnx}  \end{bsmallmatrix}, \begin{bsmallmatrix} I_{\dnx} \\ p\kron I_{\dnx}  \end{bsmallmatrix}\right). \)
Thus, by representing $L(p)$ as the linear fractional transformation $L(p) = \Delta(p)\star \bar{L} = L_{22}+L_{21}\Delta(p)(I-L_{11}\Delta(p))^{-1}L_{12}$ with $L_{11},\dots,L_{22}$ as in~\eqref{eq:LFT-matrices}, we obtain the LMI conditions of Theorem \ref{thm:stabilizing-sch-dependence} that are linear and convex in $p$. Finally, given that $\mb{P} = \mr{co}(\{\mt{p}^{i}\}_{i=1}^{n_\mathrm{v}})$, i.e., $\mb P$ is a polytope, multi-convexity of~\eqref{eq:quad-matrix-inequality-multipliers} allows to equally represent these constraints by a finite set of LMIs, specified at the vertices $\mt{p}^{i}$ of $\mathbb{P}$ \cite{ApAd98}, which concludes the proof. 
\end{proof}
{We have now derived analysis conditions in terms of matrix (in)equality constraints, which can be solved as an SDP. The conditions verify from data whether a given LPV controller $K(p_k)$ is stabilizing the closed-loop. In the next section, we will consider the controller \emph{also as an unknown} and turn the conditions of Theorem~\ref{thm:stabilizing-sch-dependence} into controller \emph{synthesis} conditions. We first want to make the following remarks} on Theorem~\ref{thm:stabilizing-sch-dependence}.

\begin{remark}[On the connections of $\mc{F}$ in Theorem~\ref{thm:stabilizing-sch-dependence}]\label{rem:mult}
    The following remarks are important to mention:
    \begin{enumerate}[label=\roman*)]
        \item {In Theorem~\ref{thm:stabilizing-sch-dependence}, to arrive at conditions that can be solved as an SDP, the decision variable $\mc{F}$ is introduced, which \emph{decouples} the multiplication of $\lyap$ with $\mc{V}$. This decoupling is essential in the formulation of the data-driven synthesis conditions that are derived in the remainder of the paper, and is one of the core contributions of this paper that enable synthesis.}
        \item Condition~\eqref{eq:coupling-condition-sch-dep} is crucial for selecting the scheduling dependency structure of both $F$ and $\lyap$. In particular, choosing the scheduling dependence of $F$ such that it matches with the cumulative dependence of the left-hand side, allows to drop the $p$-dependent terms in~\eqref{eq:coupling-condition-sch-dep} and formulate the condition on the level of the involved matrices only, making the relation scheduling \emph{in}dependent. This significantly reduces the complexity of the synthesis condition that is derived based on~\eqref{eq:ddLyap-1-general-3} and~\eqref{eq:coupling-condition-sch-dep}.
        \item Note that when we partition $\mc{F}$ and $F_Q$ in~\eqref{eq:F=VZ} as $\mc{F} = \begin{bmatrix} F_0 & \bar{F} & \bar{\bar{F}}\end{bmatrix}$ and $F_Q = \begin{bmatrix} F_{11} & F_{12} \\ F_{21} & F_{22} \end{bmatrix}$, the matrix $F_Q$ is based on just the re-shuffling of the terms of $\mathcal{F}$, i.e., $F_{11}\in\R^{\Nd-1\times\dnx}$ results from the scheduling independent term $F_0$, matrices $F_{12}\in\R^{\Nd-1\times n_\mr{x}\dnp }$ and $F_{21}\in\R^{(\Nd-1)\dnp\times n_\mr{x}}$ result from $\bar{F}$, while $F_{22}\in\R^{(\Nd-1)\dnp\times n_\mr{x}\dnp }$ is based on $\bar{\bar{F}}$. Hence, is it not necessary to define \emph{both} $\mc{F}$ and $F_Q$ as decision variables in the SDP.
    \end{enumerate}
\end{remark}

\subsection{Stabilizing controller synthesis}
We have now obtained a set of fully data-based linear constraints, which can be solved as an SDP, that allow to analyze closed-loop stability of the feedback interconnection of a given LPV controller with an unknown LPV system. 
In this section, we further extend this result by deriving
linear constraints for data-driven \emph{synthesis} of a stabilizing LPV controller. Note that in this case, the controller is a decision variable, which yields the linear conditions in Theorem~\ref{thm:stabilizing-sch-dependence} nonlinear. The following result recasts this problem, which allows to \emph{synthesize} the LPV controller using a set of linear constraints.

\begin{theorem}[Data-driven stabilizing feedback synthesis]\label{th:stabilizing-data-based} 
	Given a data-set $\dataset$, satisfying Condition~\ref{cond:rank-cond}, from a system that can be represented by~\eqref{eq:LPVSS}. For a $\lyap\in\mb{S}^{\dnx}$, let the matrices $\mc{F}\in\mb{R}^{\Nd-1\times\dnx(1+\dnp+\dnp^2)}$ and $F_Q\in\mb{R}^{(\Nd-1)(1+\dnp)\times \dnx (1+\dnp)}$ be defined as in~\eqref{eq:F=VZ}. 	
	If there exist a $\lyap \succ 0$, a $\mc{F}$, and $Y_0 \in \mathbb{R}^{\dnu\times\dnx}$, $\bar{Y} \in \mathbb{R}^{\dnu\times\dnx\dnp}$, $\Xi\in\mb{S}^{4\dnp\dnx}$ that satisfy 
	\begin{equation}\label{eq:coupling-condition-sch-dep-7}
    	\begin{bmatrix} 
    	   \lyap & 0 & 0 \\
    	   0 & I_{\dnp}\kron\lyap & 0 \\
    	   Y_0 & \bar{Y} & 0 \\
    	   0 & I_{\dnp}\kron Y_0 &  I_{\dnp}\kron\bar{Y}
    	\end{bmatrix} = \mcG \mc{F},
    \end{equation}
    for the given data dictionary and also satisfy the LMI conditions in~\eqref{eq:quad-matrix-inequality-multipliers} for all $p\in\mb{P}$, with $\Delta(p), W, L_{11},\dots,L_{22}$ as given in~\eqref{eq:LFT-matrices}, then 
	\begin{equation}\label{eq:KSD-stab}
		K_0  = Y_0 \lyap^{-1}, \quad  \bar{K} = \bar{Y} (I_{\dnp}\kron {\lyap} )^{-1},
	\end{equation}
	gives an LPV state-feedback controller $K(p)$ in terms of~\eqref{eq:controllaw:dependencyK} that 
	guarantees stability of the closed-loop interconnection %
	\eqref{eq:closedloopsys}. If $\mb{P} = \mr{co}(\{\mt{p}^{i}\}_{i=1}^{n_\mathrm{v}})$, then~\eqref{eq:quad-matrix-inequality-multipliers} with~\eqref{eq:LFT-matrices} is only required to be satisfied on $\{\mt{p}^{i}\}_{i=1}^{n_\mathrm{v}}$, corresponding to an SDP.
\end{theorem}
\begin{proof}
    The proof of this result is built on the proof of Theorem~\ref{thm:stabilizing-sch-dependence} {by recasting the matrix constraints in the analysis problem to constraints for a synthesis problem. For the synthesis case,  $K_0, \bar{K}$ in the controller structure~\eqref{eq:pfth1:1} are now  decision variables, but they only appear in the constraint~\eqref{eq:coupling-condition-sch-dep-4}.} Substituting $\mathcal{M}_\textsc{CL}$ from condition~\eqref{eq:consist-cond-general} into~\eqref{eq:coupling-condition-sch-dep-4}, yields
    \begin{equation}\label{eq:coupling-condition-sch-dep-6}
    	\begin{bmatrix}
    	{\lyap}&0&0\\0&I_{\dnp }\otimes {\lyap}&0\\
    	K_0{\lyap}&\bar{K}(I_{\dnp }\otimes {\lyap})&0\\
    	0&I_{\dnp }\otimes(K_0{\lyap})&I_{\dnp }\otimes \bar{K}(I_{\dnp }\otimes {\lyap})
    	\end{bmatrix}=\mcG \mathcal{F}.
    \end{equation}
    As~\eqref{eq:coupling-condition-sch-dep-6} is nonlinear in the decision variables $K_0$, $\bar{K}$ and $P$, introduce $Y_0=K_0 {\lyap}$ and $\bar{Y}=\bar{K}(I_{\dnp }\otimes {\lyap})$, which transforms~\eqref{eq:coupling-condition-sch-dep-6} as~\eqref{eq:coupling-condition-sch-dep-7} to arrive at a condition that is linear in the new decision variables~$Y_0, \bar{Y}, \lyap, \mc{F}$. {With the matrix equality constraint~\eqref{eq:coupling-condition-sch-dep-4} recasted as~\eqref{eq:coupling-condition-sch-dep-7}, we have by Theorem~\ref{thm:stabilizing-sch-dependence} that if there exist $\lyap \succ 0, Y_0, \bar{Y}, \mc{F}, \Xi$ that satisfy the equality constraint~\eqref{eq:coupling-condition-sch-dep-7} and the LMI conditions in~\eqref{eq:quad-matrix-inequality-multipliers} for all $p \in \mb P$ and with $\Delta(p), W, L_{11},\dots,L_{22}$ as given in~\eqref{eq:LFT-matrices}, then there exist an LPV controller~\eqref{eq:controllaw} that stabilizes the closed-loop system. The controller parametrization $K_0, \bar{K}$ can be recovered with~\eqref{eq:KSD-stab}, such that $K(p_k) = K_0 + \bar{K}(p_k\kron I_{\dnx})$. Finally, using the same arguments as in Theorem~\ref{thm:stabilizing-sch-dependence}, if $\mb{P} = \mr{co}(\{\mt{p}^{i}\}_{i=1}^{n_\mathrm{v}})$, then multi-convexity of~\eqref{eq:quad-matrix-inequality-multipliers} allows to solve the synthesis problem as a feasibility problem subject to a finite number of LMI constraints, which are defined on $\{\mt{p}^{i}\}_{i=1}^{n_\mathrm{v}}$, concluding the proof.} %
\end{proof}
Under a polytopic $\mathbb{P}$ the conditions of Theorem~\ref{th:stabilizing-data-based} are defined on the finite number of generators of $\mathbb{P}$, providing an SDP that can easily be solved using off-the-shelf methods. Hence, we obtained easily implementable approach for the \emph{direct} design of a stabilizing LPV state-feedback controller from only a single data sequence $\dataset$ collected from the unknown LPV system, provided that $\dataset$ satisfies Condition~\ref{cond:rank-cond}. 
The next section presents extensions of this synthesis approach by incorporating performance measures.

\section{Data-driven control synthesis with performance objectives} \label{s:datadrivenperf}
{Similar to Section~\ref{ss:stabanalysis}, we can establish direct data-driven performance analysis methods for a given controller $K(p_k)$ by extending existing model-based performance analysis approaches, summarized in Appendix~\ref{app:review} to the data-driven case, which, in turn, can be reformulated to synthesise $K(p_k)$ directly. Here, for the sake of compactness, we will only provide the synthesis results, as the analysis results directly follow from them.}
\subsection{Quadratic performance-based synthesis}
A controller achieves quadratic closed-loop performance for given weighting matrices $Q\succeq 0,R\succ 0$ if the controller minimizes the infinite horizon quadratic cost
\begin{equation}\label{eq:InfHorcostfunction}
	J(x,u) = {\sum_{k=0}^\infty} x_k^\top Q x_k + u_k^\top R u_k.
\end{equation}
We now extend the known conditions for model-based quadratic performance LPV controller synthesis, see, e.g.,~\cite{rotondo2015linear} and Lemma~\ref{lem:LQR-LPV-model-based} in the appendix, to the data-based setting. 
\begin{theorem}[Data-driven quadratic performance optimal synthesis]\label{th:lqrLPV-result-data-based}
    Given a data-set $\dataset$, satisfying Condition~\ref{cond:rank-cond},  from a system that can be represented by~\eqref{eq:LPVSS}. For a $\lyap\in\mb{S}^{\dnx}$, let the matrices $\mc{F}\in\mb{R}^{\Nd-1\times\dnx(1+\dnp+\dnp^2)}$ and $F_Q\in\mb{R}^{(\Nd-1)(1+\dnp)\times \dnx (1+\dnp)}$ be defined as in~\eqref{eq:F=VZ}.
	Let $\lyap$ be the minimizer of $\sup_{p\in\mb{P}}\mr{trace}(\lyap)$ among all possible choices of $\lyap \succ 0$, 
	$\Xi\in\mb{S}^{4\dnp\dnx}$, ${F}_Q$, and  $\mc{Y}:=\begin{bmatrix} Y_0 & \bar{Y} \end{bmatrix}$, such that, for all $p\in\mb{P}$, both~\eqref{eq:coupling-condition-sch-dep-7} and %
	\eqref{eq:quad-matrix-inequality-multipliers} are satisfied, where %
	\begin{subequations}\label{eq:LFT-LQR}
        \begin{align}
            W = & \begin{bmatrix} 
                \lyap_0 & (\ast)^\top  & (\ast)^\top &  (\ast)^\top \\
                \mc{\Xf}F_Q & \lyap_0 & 0 & 0 \\
                \begin{bmatrix} Q^{\frac{1}{2}}\lyap  & 0\end{bmatrix} & 0 & I_{n_\mr{x}} & 0 \\
                R^{\frac{1}{2}}\mc{Y} & 0 & 0 &  I_{n_\mr{u}}
                \end{bmatrix}, & \Delta(p) = & \diag(p)\otimes I_{2n_\mr{x}} \\
            L_{11}  = & 0_{2\dnx\dnp}, \vphantom{\Bigg)}& L_{12} = & \begin{bmatrix} \onem_{\dnp}\otimes I_{2\dnx} & 0_{2\dnx\dnp\times \dnx+\dnu} \end{bmatrix},\\
                L_{21} = & \begin{bmatrix} 
		              0_{\dnx\times 2\dnx\dnp}\\ 
		              I_{\dnp}\otimes \begin{bmatrix} I_{\dnx} & 0 \end{bmatrix}\\ 
		              0_{\dnx\times 2\dnx\dnp}\\ 
		              I_{\dnp} \otimes \begin{bmatrix} 0 & I_{\dnx} \end{bmatrix}\\ 
		              0_{\dnx+\dnu\times 2\dnx\dnp}\end{bmatrix}, &
		         L_{22}  = &\begin{bmatrix}
		              \begin{bmatrix} I_{\dnx} & 0 \end{bmatrix} & 0 \\
		              \onem_{\dnp}\otimes 0_{\dnx\times 2\dnx} & 0\\
		              \begin{bmatrix} 0 & I_{\dnx} \end{bmatrix} & 0 \\
		              \onem_{\dnp}\otimes 0_{\dnx\times 2\dnx} & 0 \\
		              0 & I_{\dnx+\dnu}
		           \end{bmatrix}.
        \end{align}
        \end{subequations}
	Then, the LPV state-feedback controller $K(p)$  as in~\eqref{eq:controllaw} with gains~\eqref{eq:KSD-stab} is a stabilizing controller for~\eqref{eq:LPVSS} 
	and achieves the minimum of $\sup_{(x,p)\in\mf{B}_{\mr{CL}}} J(x,u=K(p)x)$. If $\mb{P} = \mr{co}(\{\mt{p}^{i}\}_{i=1}^{n_\mathrm{v}})$ then~\eqref{eq:quad-matrix-inequality-multipliers} with \eqref{eq:LFT-LQR} is only required to be satisfied on $\{\mt{p}^{i}\}_{i=1}^{n_\mathrm{v}}$, corresponding to an SDP.
\end{theorem}
\begin{proof}
	Similar to the proof of Theorem~\ref{th:stabilizing-data-based}, first, the matrix inequality~\eqref{eq:LQR-LPV-model-based:a} of the model-based results of Lemma~\ref{lem:LQR-LPV-model-based} in the appendix
    is extended to its data-based counterpart using the relation $A_\mr{CL}(p)\lyap = \Xf F(p)$, where $F(p)$ satisfies~\eqref{eq:F=VZ} and~\eqref{eq:coupling-condition-sch-dep}. {This gives~\eqref{eq:coupling-condition-sch-dep-6} and
    \begin{equation}\label{eq:pf:lqr:inbetween}
        \begin{bmatrix}
			\lyap & (*)^\top & (*)^\top  & (*)^\top \\
			\Xf F(p) & \lyap & 0  & 0 \\
			Q^{\frac{1}{2}}\lyap & 0 & I_{n_\mr{x}} & 0 \\
			R^{\frac{1}{2}}K(p)\lyap & 0 & 0 & I_{n_\mr{u}} 
			\end{bmatrix}\succ 0,
    \end{equation}
    where
    \begin{equation}\label{eq:pf:lqr:truukje}
        K(p)\lyap = \begin{bmatrix}K_0 & \bar{K}\end{bmatrix}\begin{bmatrix} I_{\dnx} \\ p\kron I_{\dnx} \end{bmatrix}\lyap \overset{\eqref{kr:mult}}{=}  \begin{bmatrix}K_0\lyap & \bar{K}(I_{\dnp}\kron\lyap)\end{bmatrix}\begin{bmatrix} I_{\dnx} \\ p\kron I_{\dnx} \end{bmatrix}.
    \end{equation}
	Introduce $Y_0=K_0 {\lyap}$, $\bar{Y}=\bar{K}(I_{\dnp }\otimes {\lyap})$, and let $\mc{Y}:= \begin{bmatrix} Y_0 & \bar{Y} \end{bmatrix}$. In the first place, this gives~\eqref{eq:coupling-condition-sch-dep-7}, and secondly, through the methods in the proof of Theorem~\ref{th:stabilizing-data-based}} allows us to rewrite the resulting data-based matrix inequality in the quadratic form~\eqref{eq:quad-matrix-inequality} with~\eqref{eq:LFT-LQR} and
	\[ L(p) = \mr{blkdiag}\left( \begin{bsmallmatrix} I_{\dnx} \\ p\kron I_{\dnx}  \end{bsmallmatrix}, \begin{bsmallmatrix} I_{\dnx} \\ p\kron I_{\dnx}  \end{bsmallmatrix}, I_{\dnx}, I_{\dnu} \right). \]
	With this formulation, the $\mc{S}$-procedure in Lemma~\ref{lem:full-block-S-procedure} can be readily applied to derive the required LMI synthesis conditions. Similarly as in Theorem~\ref{th:stabilizing-data-based}, if $\mb{P} = \mr{co}(\{\mt{p}^{i}\}_{i=1}^{n_\mathrm{v}})$, multi-convexity of~\eqref{eq:quad-matrix-inequality-multipliers} allows to solve the data-driven synthesis problem as a feasibility problem, subject to a finite number of linear matrix (in)equality constraints that are defined on $\{\mt{p}^{i}\}_{i=1}^{n_\mathrm{v}}$, concluding the proof.
\end{proof}
{To turn Theorem~\ref{th:lqrLPV-result-data-based} into an analysis condition with a given $K(p_k)$, the LMI conditions can be directly obtained from the application of Lemma~\ref{lem:full-block-S-procedure} on~\eqref{eq:pf:lqr:inbetween}.}

\subsection{$\htwo$-norm performance-based synthesis}

To consider \emph{induced gains}-based performance metrics, e.g., the widely used $\mathcal{H}_2$ or $\ell_2$ performance metrics in LTI control, we will introduce a representation of general controller configurations in terms of the closed-loop \emph{generalized plant} concept. A generalized LPV plant with state measurements is given by
\begin{subequations}\label{eq:auglpvss}
\begin{align}
	x_{k+1} & = A(p_k) x_k + B(p_k) u_k +  B_\mr{w}(p_k) w_k,\\
	   y_k & =  x_k  \label{eq:auglpvss:b}\\
	   z_k & = C_\mr{z}(p_k) x_k + D_\mr{zu}(p_k) u_k + D_\mr{zw}(p_k) w_k, 
\end{align}
\end{subequations}
where $w_k\in\R^{\dnw}$ is the \emph{generalized disturbance} signal (containing, e.g., reference signals, load disturbances, etc.), $z_k\in\R^{n_\mr{w}}$ is the \emph{generalized performance} signal (containing, e.g., the tracking error or control effort), and $y_k\in\R^{\dnx}$ is the \emph{generalized control} signal, i.e., the signal that is available for the controller. To avoid complexity by using performance shaping filters, and to be compatible with the considered quadratic performance concept in~\eqref{eq:InfHorcostfunction}, consider 
\begin{equation}\label{eq:auglpvss-matrices}
	B_\mr{w}(p_k) = I_{\dnx}, \quad C_\mr{z}(p_k) = \begin{bmatrix} Q^{\frac{1}{2}} \\ 0 \end{bmatrix}, \quad D_\mr{zu}(p_k) = \begin{bmatrix} 0 \\ R^{\frac{1}{2}} \end{bmatrix}, \quad D_\mr{zw}(p_k)=0.
\end{equation}
This gives $w_k\in\R^{n_{\mr x}}$ and $z_k\in\R^{n_{\mr x}+n_{\mr u}}$. Closing the loop with feedback law~\eqref{eq:controllaw}, yields the LPV closed-loop generalized plant:
\begin{subequations}\label{eq:CLgenplant}
\begin{align}
	x_{k+1} & = \big(A(p_k) + B(p_k)K(p_k)\big) x_k +    w_k,\\
	   z_k & = \begin{bmatrix}
	   Q^{\frac{1}{2}} \\R^{\frac{1}{2}}K(p_k)
	   \end{bmatrix}x_k.
\end{align}
\end{subequations}
We can characterize the performance of $w \rightarrow z$ in terms of the so-called $\htwo$-norm\footnote{This is an LPV-extension of the classical $\mathcal{H}_2$-norm for LTI systems. There are multiple formulations, but we take here the definition from~\cite{deCaigny2010_H2LPV}.}, which is defined for an exponentially stable~\eqref{eq:CLgenplant} as
\begin{equation}\label{eq:h2normdef}
	\|\Sigma\|_{\htwo}:= \left(\limsup_{N\to\infty}\mb{E}\left\{\tfrac{1}{N}{\textstyle\sum_{k=0}^N} z_k^\top z_k\right\}\right)^{\tfrac{1}{2}},
\end{equation}
given that $w_k$ is a white noise signal. Here $\mb{E}$ denotes the expectation w.r.t. $w$. We can now formulate the data-based analog of Lemma~\ref{lem:H2-LPV-conditions-model-based}, \cite[Lem.~1]{deCaigny2010_H2LPV},
which synthesizes a controller that guarantees closed-loop stability and performance in terms a bound~$\gamma$ on the $\htwo$-norm of~\eqref{eq:CLgenplant}.
\begin{theorem}[Data-driven $\htwo$-norm performance synthesis]\label{th:H2-LPV-result-data-based}
	Given a performance objective $\gamma>0$, and a data-set $\dataset$ satisfying Condition~\ref{cond:rank-cond} from the system represented by~\eqref{eq:LPVSS}. If there exist matrices $\lyap\in\mb{S}^{\dnx}$ and $S\in\mb{S}^{\dnu}$, $\Xi\in\mb{S}^{4\dnp\dnx}$, ${F}_Q$ as in~\eqref{eq:F=VZ}, and  $\mc{Y}:=\begin{bmatrix} Y_0 & \bar{Y} \end{bmatrix}$, such that~\eqref{eq:coupling-condition-sch-dep-7},
	\eqref{eq:quad-matrix-inequality-multipliers} and
	\begin{equation}\label{eq:H2-LPV-conditions-data-based}
    	\begin{bmatrix}
    	S & R^{\frac{1}{2}}\mc{Y}\begin{bmatrix}
    	I_{n_\mr{x}}\\p\otimes I_{n_\mr{x}} 
    	\end{bmatrix}\\
    	(*)^\top & {\lyap}_0
    	\end{bmatrix}\succ 0, \qquad \lyap - I_{n_\mr{x}} \succ  0,
    	\qquad
    	\mr{trace}(Q{\lyap})+\mr{trace}(S)  \prec \gamtwo^2,
	\end{equation}
	are satisfied for all $p\in\mathbb{P}$, where in~\eqref{eq:quad-matrix-inequality-multipliers},
	\begin{equation}\label{eq:h2syn_Wdef}
	    W  = \begin{bmatrix}
	        \lyap_0-I_{n_\mr{x0}} & \mc{\Xf}{F}_Q \\
	        (*)^\top & \lyap_0
	    \end{bmatrix}, \qquad  I_{n_\mr{x0}}=\mr{blkdiag}(I_{n_\mr{x}},0_{\dnx\dnp}),
	\end{equation}
	and $P_0,\mc{\Xf}, \Delta(p), L_{11}, \dots, L_{22}$ are as in~\eqref{eq:LFT-matrices}, then, the resulting LPV state-feedback controller $K(p)$ in~\eqref{eq:controllaw} with gains~\eqref{eq:KSD-stab} is a stabilizing controller for~\eqref{eq:LPVSS} and achieves 
	an $\htwo$-norm of  the closed-loop system~\eqref{eq:CLgenplant} that is less than $\gamtwo$. If $\mb{P} = \mr{co}(\{\mt{p}^{i}\}_{i=1}^{n_\mathrm{v}})$ then~\eqref{eq:H2-LPV-conditions-data-based} and \eqref{eq:quad-matrix-inequality-multipliers} with \eqref{eq:LFT-matrices} are only required to be satisfied on $\{\mt{p}^{i}\}_{i=1}^{n_\mathrm{v}}$, corresponding to an SDP.
\end{theorem}
\begin{proof}
	The proof follows the same lines as for Theorem~\ref{th:stabilizing-data-based} and Theorem~\ref{th:lqrLPV-result-data-based}. {The data-based counterparts of the matrix inequalities~\eqref{eq:H2-LPV-conditions-model-based} for the model-based synthesis conditions of Lemma~\ref{lem:H2-LPV-conditions-model-based} in the appendix
    are obtained using the relationship $A_\mr{CL}(p)\lyap = \Xf F(p)$ and~\eqref{eq:pf:lqr:truukje}. Writing the LMIs (quadratic in $p$) in the}
	quadratic form~\eqref{eq:quad-matrix-inequality}, with $W$ as in~\eqref{eq:h2syn_Wdef} and $L(p) = \mr{blkdiag}\left( \begin{bsmallmatrix} I_{\dnx} \\ p\kron I_{\dnx}  \end{bsmallmatrix}, \begin{bsmallmatrix} I_{\dnx} \\ p\kron I_{\dnx}  \end{bsmallmatrix}\right)$, allows to apply the full-block $\mc{S}$-procedure. The remainder of the proof follows directly from Theorem~\ref{th:stabilizing-data-based}.
\end{proof}
{To turn Theorem~\ref{th:H2-LPV-result-data-based} into an analysis condition for the case when~$K(p)$ is given, the LMI conditions directly follow from Theorem~\ref{th:H2-LPV-result-data-based}.}

\subsection{$\ltwo$-gain performance-based synthesis}\label{ss:ltwosyn}
Another widely used performance metric is the (induced) $\ltwo$-gain of a system, which can be seen as the generalization of the $\mathcal{H}_\infty$ norm for LTI systems to the LPV case and is a widely used performance metric in model-based LPV control. The induced $\ltwo$-gain of a system is defined as the infimum of $\gamma>0$ such that for all trajectories in $\mf{B}$, with $x_0=0$, we have \( \|z\|_2\le\gamma\|w\|_2, \)
where $\|\cdot\|_2$ denotes the $\ltwo$-norm of a signal, cf.~\cite{Schaft2017_L2book}. Following the well-known result in~\cite{GaAp94} on $\ltwo$-gain LPV state-feedback synthesis, see also Lemma~\ref{lem:Hinf-LPV-conditions-model-based}, we now formulate a fully data-based method to synthesize a state-feedback controller $K(p_k)$ that guarantees stability and an $\ltwo$-gain less than $\gamma$ for the closed-loop system~\eqref{eq:CLgenplant}.
\begin{theorem}[Data-driven $\ltwo$-gain performance synthesis]\label{th:Hinf-LPV-result-data-based}
	Given a performance objective $\gamma>0$, and a data-set $\dataset$ satisfying Condition~\ref{cond:rank-cond} from the system represented by~\eqref{eq:LPVSS}. If there exist matrices  $\lyap\in\mb{S}^{\dnx}$ with $\lyap\succ 0$ and $\Xi\in\mb{S}^{4\dnp\dnx}$, ${F}_Q$ as in~\eqref{eq:F=VZ}, and  $\mc{Y}:=\begin{bmatrix} Y_0 & \bar{Y} \end{bmatrix}$, such that~\eqref{eq:coupling-condition-sch-dep-7} and~\eqref{eq:quad-matrix-inequality-multipliers} are satisfied for all $p\in\mathbb{P}$, where in~\eqref{eq:quad-matrix-inequality-multipliers}: 
	\begin{subequations}\label{eq:LFT-L2}
	\begin{align}
	    W & = \begin{bmatrix}
		\lyap_0 & (*)^\top & (*)^\top & (*)^\top & 0 \\
		\mc{\Xf}F_Q & \lyap_0 & 0 & 0 & (*)^\top \\
		 \begin{bmatrix}  Q^{\frac{1}{2}}\lyap  & 0\end{bmatrix} & 0 & \gaminf I_{n_\mr{x}} & 0 & 0 \\
		 R^{\frac{1}{2}}\mc{Y} & 0 & 0 &  \gaminf I_{n_\mr{u}} & 0 \\
		 0 & \begin{bmatrix} I_{n_\mr{x}} & 0 \end{bmatrix} & 0 & 0 & 
		 \gaminf I_{n_\mr{x}}
            \end{bmatrix}, & \Delta(p) & = \diag(p)\otimes I_{2n_\mr{x}} \\
        L_{11} & =  0_{2\dnx\dnp}, & L_{12} & = \begin{bmatrix} \onem_{\dnp}\otimes I_{2\dnx} & 0_{2\dnx\dnp\times2\dnx+\dnu}
		\end{bmatrix}, \vphantom{\Big)}\\
        L_{21} & =  \begin{bmatrix} 0_{\dnx\times 2\dnx\dnp} \\ I_{\dnp}\otimes \begin{bmatrix} I_{\dnx} & 0 \end{bmatrix} \\ 
            0_{\dnx\times 2\dnx\dnp} \\ I_{\dnp}\kron\begin{bmatrix} 0 & I_{\dnx} \end{bmatrix}\\ 
            0_{2\dnx+\dnu\times2\dnx\dnp}
		\end{bmatrix}, &  L_{22} & =  \begin{bmatrix} \begin{bmatrix} I_{\dnx} & 0 \end{bmatrix} & 0 \\
		 \onem_{\dnp}\kron 0_{n_\mr{x}\times 2n_\mr{x}}&0\\
		\begin{bmatrix} 0 & I_{\dnx} \end{bmatrix} &0\\
		 \onem_{\dnp}\kron 0_{n_\mr{x}\times 2n_\mr{x}} & 0\\
		 0 & I_{2\dnx+\dnu}
		\end{bmatrix},
	\end{align}
	\end{subequations}
    then the resulting LPV state-feedback controller $K(p)$  as in~\eqref{eq:controllaw} with gains~\eqref{eq:KSD-stab} is a stabilizing controller for~\eqref{eq:LPVSS} and achieves 
	an $\ltwo$-gain of the closed-loop sytem~\eqref{eq:CLgenplant} that is less than $\gamtwo$. If $\mb{P} = \mr{co}(\{\mt{p}^{i}\}_{i=1}^{n_\mathrm{v}})$, then~\eqref{eq:quad-matrix-inequality-multipliers} with \eqref{eq:LFT-L2} is only required to be satisfied on $\{\mt{p}^{i}\}_{i=1}^{n_\mathrm{v}}$, corresponding to an SDP.
\end{theorem}
\begin{proof}
    The proof follows the same lines as for Theorems~\ref{th:stabilizing-data-based}--\ref{th:H2-LPV-result-data-based}, i.e., substitution of $A_\mr{CL}(p)\lyap = \Xf F(p)$ in~\eqref{eq:Hinf-LPV-conditions-model-based} and manipulations with~\eqref{eq:pf:lqr:truukje}. The quadratic form~\eqref{eq:quad-matrix-inequality} is obtained with $W$ in~\eqref{eq:LFT-L2} and $L(p):=\mr{blkdiag}\left( \begin{bsmallmatrix} I_{\dnx} \\ p\kron I_{\dnx}  \end{bsmallmatrix}, \begin{bsmallmatrix} I_{\dnx} \\ p\kron I_{\dnx}  \end{bsmallmatrix}, I_{\dnx}, I_{\dnu}, I_{\dnx} \right)$, from which $L_{11},\dots,L_{22}$ in~\eqref{eq:LFT-L2} are derived. The remainder of the proof directly follows from Theorems~\ref{th:stabilizing-data-based}--\ref{th:H2-LPV-result-data-based}.
\end{proof}
We want to emphasize here, that if $\mb{P} = \mr{co}(\{\mt{p}^{i}\}_{i=1}^{n_\mathrm{v}})$, the LMI conditions of the data-driven synthesis methods discussed in this paper are defined on the finite set of generators $\mt{p}^{i}$. Hence, they can be solved as an SDP using off-the-shelf tools and they provide LPV controllers that stabilize the unknown LPV system for all $p\in\mb{P}$, purely based on $\dataset$. 
{Similar to Theorems~\ref{th:stabilizing-data-based}--\ref{th:H2-LPV-result-data-based}, the data-based conditions for the $\ltwo$-gain analysis problem under a given controller can be directly derived from Theorem~\ref{th:Hinf-LPV-result-data-based}.}
\begin{remark}[Maximizing $\htwo$/$\ltwo$ performance during synthesis]
	A common practice in $\htwo$-norm/$\ltwo$-gain-based synthesis algorithms is to synthesize controllers that guarantee the smallest upper bound on the true $\htwo$-norm/$\ltwo$-gain of the closed-loop. The results in Theorem~\ref{th:H2-LPV-result-data-based} and Theorem~\ref{th:Hinf-LPV-result-data-based} allow for formulating the SDPs with the minimization of $\gamma$ as well, because $\gamma$ appears linearly in the LMI conditions.
\end{remark}
\begin{remark}[Robust controller design]\label{rem:partitioning-of-F-towards-robust}
    Our results consider analysis and control synthesis problems where $A,B,K$ are affinely dependent on $p$. This structural dependency in the closed-loop system is clearly revealed in our derivations for the synthesis methods by choosing the partitioning of $F_Q$ as in Remark~\ref{rem:mult}-(iii). Hence, $F_{11}$ represents the terms {in the closed-loop} that are \emph{independent} of the scheduling,  $F_{12},F_{21}$ represent the affine dependence on $p$ in the closed-loop and $F_{22}$ represents the quadratic dependence, emerging from the multiplication of $B(p)$ and $K(p)$. This clear distinction allows us to have control over {whether the controller $K$ should be scheduling dependent or independent.} More specifically, if we enforce $F_{22}=0$, the synthesized controller becomes scheduling \emph{independent}, which results in the synthesis of a \emph{robust} state-feedback controller. {How this special case of our synthesis method connects to the existing robust LTI data-driven synthesis methods of, e.g.,~\cite{berberich2022combining}, is an open question.}
\end{remark}
\begin{remark}[Control synthesis for nonlinear systems]
    {So far, we have considered LPV systems where the scheduling is a measurable signal that varies independently from the state or inputs. When the LPV framework is used in practice, however, the LPV description often acts as a surrogate of a nonlinear system, meaning that the scheduling signal~$p$ is defined by a \emph{scheduling-map} $\psi:\mb{X}\times\mb{U}\to\mb{P}$ that schedules the nonlinearities and time-variations of the underlying system. If the Condition~\ref{cond:rank-cond} is satisfied, then the resulting controller provides closed-loop guarantees for the underlying nonlinear system in terms of the LPV embedding principle. For details and how to ensure stability and performance guarantees for all forced equilibria, see~\cite{verhoek2023directgen}. For alternative direct data-driven state-feedback control methods for nonlinear systems, see, e.g.,~\cite{hu2023learning, dePersis2023annualrev, dePersis2023learningNLcancel}.}
\end{remark}

\section{Handling noisy measurement data}\label{s:noise}

In this section, we consider the situation where the measurements in $\dataset$ are noisy.
Note that recently, data-driven LPV control methods have been proposed in the literature that can handle noisy data, e.g.,~\cite{MillerSznaier2022, verhoek2024decoupling}. {These methods, however, utilize a deterministic set-membership argument, which involves constructing a set that describes all the possible systems that could have generated the data, given a bound on the disturbance. With a similar argument, robust data-driven analysis and control formulations have been developed for LTI systems, see, e.g., the methods from~\cite{koch2020verifying, koch2021provably, vanWaarde2022_matrixS} for analysis and~\cite{berberich2022combining, berberich2020robust, vanwaarde2023quadratic} for controller design (note that these methods synthesize LTI controllers). These methods, however, are focused on handling process disturbances, which corresponds to an ARX setting in system identification. The method that we present in this section is focused on \emph{measurement noise}, corresponding to an \emph{output-error} (OE) setting in system identification, i.e., the noise does \emph{not} propagate through the system dynamics, contrary to~\cite{berberich2020robust, MillerSznaier2022, verhoek2024decoupling}. Note that in system identification, the OE setting is considered to be more realistic to real-world applications. Our considered setting} can be seen as the LPV extension of~\cite[Sec.~V.A]{dePersisTesi2020}. Moreover, in our setting, we do not consider a particular bound or statistical property on the noise itself, only a \emph{signal-to-noise ratio} (SNR)-like condition with respect to the data-dictionary, which is general enough to include a wide range of measurement noise settings. 

\subsection{Setting}
Consider the system~\eqref{eq:LPVSS}, but now suppose that our state observations are corrupted by a possibly colored measurement noise term $\varepsilon$, i.e.,
\begin{equation}
    z_k = x_k + \varepsilon_k, 
\end{equation}
where we assume $\varepsilon_k$ to be independent w.r.t. $x_k$. 
The goal is now to design a stabilizing controller using the measurements $z_k$, i.e., our data-dictionary is given as
\begin{equation}
    \dataset^{\varepsilon} \coloneq \left\{u_k^\mr{d}, p_k^\mr{d}, z_k^\mr{d} \right\}_{k=1}^\Nd.
\end{equation}
Let us collect the noise samples associated to the measurements in $\dataset^\varepsilon$ in the matrices
\begin{equation}\label{eq:noisetrajectories}
    \Em :=\begin{bmatrix} \varepsilon^\mr{d}_1 & \dots & \varepsilon^\mr{d}_{\Nd-1} \end{bmatrix}, \qquad \Ef :=\begin{bmatrix} \varepsilon^\mr{d}_2 & \dots & \varepsilon^\mr{d}_{\Nd} \end{bmatrix},
\end{equation}
such that the noisy state-observations in $\dataset^{\varepsilon}$ are collected as
\begin{equation}
    \Zm \coloneq X + \Em, \qquad \Zf \coloneq \Xf + \Ef.
\end{equation}
Similar to~\eqref{eq:data-matrices}, we define 
\[ \Zm^\mt{p}:=\begin{bmatrix} p^\mr{d}_1\otimes z^\mr{d}_1 & \dots & p^\mr{d}_{\Nd-1}\otimes z^\mr{d}_{\Nd-1} \end{bmatrix}, \ \text{ and } \ \Em^\mt{p}:=\begin{bmatrix} p^\mr{d}_1\otimes\varepsilon^\mr{d}_1 & \dots & p^\mr{d}_{\Nd-1}\otimes\varepsilon^\mr{d}_{\Nd-1} \end{bmatrix}.\] 
Note that $\Zm^\mt{p} = \Xp + \Em^\mt{p}$. 
We again consider a persistence of excitation-like condition, similar to Condition~\ref{cond:rank-cond}:
\begin{condition}\label{cond:noisepe}
    Both the matrices $\begin{bmatrix} \Zm^\top & (\Zm^\mt{p})^\top & U^\top & (\Up)^\top \end{bmatrix}^\top$ and $\Zf$
    have full row rank.
\end{condition}
Additionally, we assume that we are given a statistical bound on the noise power:
\begin{assumption}\label{ass:noisesnr}
    For some $\epsilon>0$, it holds that $\tfrac{1}{\Nd}R_-R^\top_- \negsemidef \tfrac{1}{\Nd}\epsilon\Zf\Zf^\top$, where $R_-:=\mc{A}\begin{bmatrix} \Em \\ \Em^\mt{p} \end{bmatrix} - \Ef$.
\end{assumption}
In \cite[Sec.~V.A]{dePersisTesi2020}, the discussion that follows the LTI variants of these assumptions is also applicable to our LPV counterparts. That is, it is easy to observe that Condition~\ref{cond:noisepe} may in fact be satisfied with an overwhelming probability for a wide range of noise processes that generate $\varepsilon_k$, e.g. if $\varepsilon_k$ is a white noise with non-zero variance. Note that $R_-$ is proportional to the additive noise perturbation on the state evolution, i.e.,
\[ R_- = \mc{A}\begin{bmatrix} \Zm \\ \Zm^\mt{p} \end{bmatrix} - Z_+  + \mc{B} \begin{bmatrix} U \\ \Up \end{bmatrix}. \]
With Assumption~\ref{ass:noisesnr}, we assume a bound on this perturbation, which reflects a bound on the \emph{power} of the noise, quantified with $\epsilon$. Hence, $\epsilon$ can be seen as an indication of the noise level in a SNR-like sense. Note that if Condition~\ref{cond:noisepe} is satisfied, there always exists a large enough $\epsilon$ such that Assumption~\ref{ass:noisesnr} holds.

\subsection{Stabilizing controller synthesis with noisy data}
We are now ready to present the direct data-driven synthesis method for noisy data. We consider here only the synthesis of a \emph{stabilizing} LPV controller. The extension of this result for the aforementioned performance metrics follows directly with minor modifications.

Intuitively, it is not possible to find a stabilizing LPV controller for arbitrary $\epsilon$, i.e., arbitrary SNR. The following result shows that, similar to a performance bound, a stabilizing feedback controller can only be found if the LMI conditions remain feasible under a specified upper bound on $\varepsilon$. %
\begin{theorem}[Data-driven stabilizing synthesis from noisy data]\label{thm:noise}
    Given $\dataset^\varepsilon$ such that Condition~\ref{cond:noisepe} and Assumption~\ref{ass:noisesnr} hold. Let $\alpha>0$ be given such that $\frac{\alpha^2}{4+2\alpha}>\epsilon$. If there exist a matrix $\lyap\in\mb{S}^{\dnx}$, $F_Q\in\mb{R}^{(\Nd-1)(1+\dnp)\times \dnx (1+\dnp)}$, $Y_0 \in \mathbb{R}^{\dnu\times\dnx}$, and $\bar{Y} \in \mathbb{R}^{\dnu\times\dnx\dnp}$ that satisfy $\lyap\posdef0$ and
    \begin{gather}
        \begin{bmatrix} 
    	   \lyap & 0 & 0 \\
    	   0 & I_{\dnp}\kron\lyap & 0 \\
    	   Y_0 & \bar{Y} & 0 \\
    	   0 & I_{\dnp}\kron Y_0 &  I_{\dnp}\kron\bar{Y}
    	\end{bmatrix} = \begin{bmatrix} \Zm \\ \Zm^\mt{p} \\ U \\ \Up \end{bmatrix} \mc{F}, \label{eq:noisethmequal}\\
    	\begin{bmatrix}
    	    \lyap-\alpha\Zf\Zf^\top \ & \ \Zf F(p) \\ (\Zf F(p))^\top & \lyap
    	\end{bmatrix} \posdef 0, \qquad \begin{bmatrix} I_{\Nd-1} & F(p) \\ F^\top(p) & \lyap \end{bmatrix} \posdef 0,\label{eq:noisethmlmis}
    \end{gather}
    for all $p\in\mb{P}$, where $F(p),\mc{F}$ are constructed with $F_Q$ as in~\eqref{eq:F=VZ}, then the LPV state-feedback controller $K(p)$ as in~\eqref{eq:controllaw} with gains~\eqref{eq:KSD-stab} is a stabilizing controller for~\eqref{eq:LPVSS}.
\end{theorem}
\begin{proof}
    The concept of this proof is based on~\cite[Thm.~5]{dePersisTesi2020}. Note that the matrix equality condition in~\eqref{eq:noisethmequal} follows directly from the proof of Theorem~\ref{th:stabilizing-data-based}.     
    We first derive a data-based representation of the closed-loop LPV system and the noise trajectories~\eqref{eq:noisetrajectories}, parametrized by $\mcV$. Recall $\mc{M}$ and $\mc{M}_\textsc{cl}$ from (the proof of) Theorem~\ref{th:closed-loop-data-based-general}. We start off with~\eqref{eq:proof1}:
    \begin{multline}\label{eq:pfth7:M}
        \mc{M} = \begin{bmatrix} \mc{A} & \mc{B} \end{bmatrix} \mc{M}_\textsc{cl} = \begin{bmatrix} \mc{A} & \mc{B} \end{bmatrix} \begin{bmatrix} \Zm \\ \Zm^\mt{p} \\ U \\ \Up \end{bmatrix}\mcV = \begin{bmatrix} \mc{A} & \mc{B} \end{bmatrix}\mcG\mcV + \begin{bmatrix} \mc{A} & \mc{B} \end{bmatrix}\begin{bmatrix} \Em \\ \Em^\mt{p} \\ 0 \\ 0 \end{bmatrix}\mcV \\ = \left(\Xf + \mc{A}\begin{bmatrix} \Em \\ \Em^\mt{p} \end{bmatrix}\right)\mcV = (\Zf + R_-)\mcV,
    \end{multline}
    where $\mcV$ also needs to satisfy
    \[ \underbrace{\begin{bmatrix}
		I_{\dnx} & 0& 0\\
		0& I_{\dnp}\otimes I_{\dnx}& 0\\
		K_0 & \bar{K} & 0\\
		0 & I_{\dnp}\otimes K_0 &  I_{\dnp}\otimes \bar{K} 
		\end{bmatrix}}_{\mc{M}_\textsc{cl}} = \begin{bmatrix}
		\Zm \\ \Zm^\mt{p} \\ U\\ \Up
		\end{bmatrix} \mcV. \]
	Now that we have a parametrization of $\mc{M}$ in terms of the measured data, the noise (by means of $R_-$) and $\mcG$, so we can substitute~\eqref{eq:pfth7:M} into~\eqref{eq:Lyapunov-LPV-model-based-2}:
	\begin{multline}\label{eq:noiseLMIworkout}
	   	\begin{bmatrix}
    	    \lyap & \lyap \VP^\top(p)(\Zf + R_-)^\top \\ (\Zf + R_-)\VP(p)\lyap  & \lyap
	   \end{bmatrix} \posdef 0 \iff  (\Zf + R_-)\VP(p)\lyap \VP^\top(p)(\Zf + R_-)^\top - \lyap = \\
	   =  \Zf \VP(p)\lyap \VP^\top(p)\Zf^\top + R_-\VP(p)\lyap \VP^\top(p)R_-^\top + \Zf \VP(p)\lyap \VP^\top(p)R_-^\top + R_-\VP(p)\lyap \VP^\top(p)\Zf^\top - \lyap \negdef 0.
	\end{multline}
	We now use a special case of Young's relation~\cite{duan2013lmis}, given by
	\[ \text{Let }S\posdef 0, \delta>0, \qquad  X^\top S Y + Y^\top S X \negsemidef \delta X^\top S X + \delta^{-1}Y^\top S Y.  \]
	Applying this to~\eqref{eq:noiseLMIworkout} with $X = (\Zf \VP(p))^\top$, $S=\lyap$ and $Y = (R_-\VP(p))^\top$, we obtain
	\begin{equation}\label{eq:pfth7:bound}
	    \text{\eqref{eq:noiseLMIworkout}}\negsemidef \underbrace{(1+\delta)\Zf \VP(p)\lyap \VP^\top(p)\Zf^\top + (1+\delta^{-1})R_-\VP(p)\lyap \VP^\top(p)R_-^\top - \lyap}_{=\Theta_0(p,\delta)}.
	\end{equation}
	Recall from~\eqref{eq:F=VZ} that $F(p)= \VP(p)\lyap$. 
    Suppose we have found a solution to~\eqref{eq:noisethmlmis}. Working out the matrix inequalities in~\eqref{eq:noisethmlmis} and applying the Schur complement gives, cf.~\eqref{eq:XarrowF(p)-general-quadratic},
	\begin{subequations}\label{eq:schuredlmis}
	\begin{align}
	    \Theta_1(p,\alpha)&=  \Zf F(p)\lyap^{-1}F^\top(p)\Zf^\top + \alpha \Zf\Zf^\top - \lyap \negdef 0, \\ 
	    \Theta_2(p) & =  F(p)\lyap^{-1}F^\top(p) - I_{\Nd-1} \negdef 0.
	\end{align}
   	\end{subequations}
	Note that $\Theta_2(p)$ is equivalent to $\VP(p)\lyap \VP^\top(p)-I_{\Nd-1}\negdef0$. As a consequence, we can bound $\Theta_0(p,\delta)$ by
	\begin{equation}\label{eq:lmiproofnoise2}
	    \Theta_0(p,\delta) \negdef \Zf \VP(p)\lyap \VP^\top(p)\Zf^\top + \delta \Zf\Zf^\top + (1+\delta^{-1})R_-R_-^\top - \lyap.
	\end{equation}
	Finally, we observe that the right-hand side of~\eqref{eq:lmiproofnoise2} is equivalent to
	\( \Theta_1(p,\alpha) + (\delta-\alpha) \Zf\Zf^\top + (1+\delta^{-1})R_-R_-^\top \), i.e., 
	\[ \Theta_0(p,\delta)\negdef (\delta-\alpha) \Zf\Zf^\top + (1+\delta^{-1})R_-R_-^\top.\]
	Hence, solutions to~\eqref{eq:noisethmlmis} provide a stabilizing LPV controller if 
	\[ (\delta-\alpha) \Zf\Zf^\top + (1+\delta^{-1})R_-R_-^\top \negdef 0 \iff R_-R_-^\top \negdef \frac{\alpha-\delta}{1+\delta^{-1}}\Zf\Zf^\top. \]
	For $\delta=\tfrac{\alpha}{2}$, this is guaranteed when $\frac{\alpha^2}{4+2\alpha}>\epsilon$, which concludes the proof.
\end{proof}
The matrix (in)equalities of Theorem~\ref{thm:noise} are quadratic in $p$. To formulate the problem in terms of a finite set of LMI constraints, we employ the full-block $\mc{S}$-procedure of Lemma~\ref{lem:full-block-S-procedure} as with the earlier synthesis problems.
\begin{proposition}[Convex synthesis problem for noisy data]\label{prop:noiselmi}
    Given $\dataset^\varepsilon$ such that Condition~\ref{cond:noisepe} and Assumption~\ref{ass:noisesnr} hold. Let $\alpha>0$ be given such that $\frac{\alpha^2}{4+2\alpha}>\epsilon$. The matrix (in)equalities~\eqref{eq:noisethmequal},\eqref{eq:noisethmlmis} hold if and only if there exist an  an $F_Q$, a $\lyap\posdef 0$, matrices $Y_0$ and $\bar{Y}$, and a multiplier $\Xi\in\mb{S}^{2\dnp(3\dnx + \Nd-1)}$ that satisfy~\eqref{eq:noisethmequal} and the LMI conditions in~\eqref{eq:quad-matrix-inequality-multipliers}, where $\Delta(p)$, $W$ and $L_{11},\dots,L_{22}$ are given by:
    \begin{subequations}\label{eq:LFT-noise}
    \begin{align}
        \Delta(p)  &= \mr{blkdiag}\big(\mr{diag}(p)\kron I_{\dnx}, \mr{diag}(p)\kron I_{\dnx}, \mr{diag}(p)\kron I_{\Nd-1}, \mr{diag}(p)\kron I_{\dnx}\big), \\
        W  &= \left[\begin{array}{c:c}
                    \begin{array}{c|c} \begin{matrix} \lyap -\alpha\Zf\Zf^\top & 0 \\ 0 & 0 \end{matrix} & \mcZf F_Q \\\hline (\mcZf F_Q)^\top & \begin{matrix} \lyap \ & \ 0 \\ 0 \ & \ 0 \end{matrix} \end{array} & \varnothing \\\hdashline \varnothing & \begin{array}{c|c} \begin{matrix} I_{\Nd-1} & 0 \\ 0 & 0 \end{matrix} &  F_Q \\\hline F_Q^\top & \begin{matrix}\ \lyap \ & \ 0 \ \\ \ 0 \ & \ 0 \ \end{matrix} \end{array} \end{array}\right], \label{eq:propnoise:WL}\\
                    L_{11}&= 0_{\dnp(3\dnx+\Nd-1)},\\
                    L_{12}&= \mr{blkdiag}\big(\onem_{\dnp}\kron I_{\dnx},\onem_{\dnp}\kron I_{\dnx},\onem_{\dnp}\kron I_{\Nd-1},\onem_{\dnp}\kron I_{\dnx} \big),\vphantom{\Big)}\\
                    L_{21}&= \mr{blkdiag}\big(\!\begin{bmatrix} 0_{\dnx\times\dnx\dnp} \\ I_{\dnx\dnp} \end{bmatrix}\!, \begin{bmatrix} 0_{\dnx\times\dnx\dnp} \\ I_{\dnx\dnp} \end{bmatrix}\!, \begin{bmatrix} 0_{\Nd-1\times(\Nd-1)\dnp} \\ I_{(\Nd-1)\dnp} \end{bmatrix}\!, \begin{bmatrix} 0_{\dnx\times\dnx\dnp} \\ I_{\dnx\dnp} \end{bmatrix}\! \big),\\
                    L_{22}&= \mr{blkdiag}\big(\!\begin{bmatrix} I_{\dnx} \\ 0_{\dnx\dnp\times\dnx} \end{bmatrix}\!,\begin{bmatrix} I_{\dnx} \\ 0_{\dnx\dnp\times\dnx} \end{bmatrix}\!,\begin{bmatrix} I_{\Nd-1} \\ 0_{(\Nd-1)\dnp\times\Nd-1} \end{bmatrix}\!,\begin{bmatrix} I_{\dnx} \\ 0_{\dnx\dnp\times\dnx} \end{bmatrix}\! \big),\vphantom{\Bigg)} \\
                    \mcZf &= \mr{blkdiag}(\Zf,I_{\dnp}\kron\Zf).
    \end{align}
    \end{subequations}
    If $\mb{P} = \mr{co}(\{\mt{p}^{i}\}_{i=1}^{n_\mathrm{v}})$ then~\eqref{eq:quad-matrix-inequality-multipliers} with \eqref{eq:LFT-noise} is only required to be satisfied on $\{\mt{p}^{i}\}_{i=1}^{n_\mathrm{v}}$, corresponding to an SDP.
\end{proposition}
\begin{proof}
First, we use the relationship in~\eqref{eq:XarrowF(p)-general-quadratic} and~\eqref{eq:F=VZ} to write~\eqref{eq:noisethmlmis} as
\begin{gather}
    \left[\begin{array}{c|c}
		\begin{matrix} I_{\dnx} \\ p\kron I_{\dnx} \end{matrix} & 0 \\ \hline  0 & \begin{matrix} I_{\dnx} \\ p\kron I_{\dnx} \end{matrix} \end{array}\right]^\top \left[\begin{array}{c|c} \begin{matrix} \lyap -\alpha\Zf\Zf^\top & 0 \\ 0 & 0 \end{matrix} & \mcZf F_Q \\\hline (\mcZf F_Q)^\top & \begin{matrix} \lyap \ & \ 0 \\ 0 \ & \ 0 \end{matrix} \end{array}\right]\left[\begin{array}{c|c}
		\begin{matrix} I_{\dnx} \\ p\kron I_{\dnx} \end{matrix} & 0 \\ \hline  0 & \begin{matrix} I_{\dnx} \\ p\kron I_{\dnx} \end{matrix} \end{array}\right] \posdef 0, \\
		\left[\begin{array}{c|c}
		\begin{matrix} I_{\Nd-1} \\ p\kron I_{\Nd-1} \end{matrix} & 0 \\ \hline  0 & \begin{matrix} I_{\dnx} \\ p\kron I_{\dnx} \end{matrix} \end{array}\right]^\top \left[\begin{array}{c|c} \begin{matrix} I_{\Nd-1} & 0 \\ 0 & 0 \end{matrix} &  F_Q \\\hline F_Q^\top & \begin{matrix} \lyap \ & \ 0 \\ 0 \ & \ 0 \end{matrix} \end{array}\right]\left[\begin{array}{c|c}
		\begin{matrix} I_{\Nd-1} \\ p\kron I_{\Nd-1} \end{matrix} & 0 \\ \hline  0 & \begin{matrix} I_{\dnx} \\ p\kron I_{\dnx} \end{matrix} \end{array}\right] \posdef 0.
\end{gather}
Combining the above two matrix inequalities allows to write them in the form of~\eqref{eq:quad-matrix-inequality} with $W$ as in~\eqref{eq:propnoise:WL} and 
\begin{equation}
    L(p):= \mr{blkdiag}\big( \begin{bmatrix} I_{\dnx} \\ p\kron I_{\dnx} \end{bmatrix}, \begin{bmatrix} I_{\dnx} \\ p\kron I_{\dnx} \end{bmatrix}, \begin{bmatrix} I_{\Nd-1} \\ p\kron I_{\Nd-1} \end{bmatrix}, \begin{bmatrix} I_{\dnx} \\ p\kron I_{\dnx} \end{bmatrix} \big).
\end{equation}
$L(p)$ can be written in LFR form with $L_{11},\dots,L_{22}$ as given in~\eqref{eq:quad-matrix-inequality}. Application of Lemma~\ref{lem:full-block-S-procedure} gives LMIs convex in $p$, which need to hold for all $p\in\mb{P}$. Finally, given that $\mb{P} = \mr{co}(\{\mt{p}^{i}\}_{i=1}^{n_\mathrm{v}})$, multi-convexity of~\eqref{eq:quad-matrix-inequality-multipliers} allows to equally represent these constraints by a \emph{finite} set of LMIs, defined at the vertices $\mt{p}^{i}$ of $\mathbb{P}$ \cite{ApAd98}, which concludes the proof.
\end{proof}

We have now presented a direct data-driven control method that allows to synthesize LPV controllers directly from noisy measurement data. Moreover, the extension of this result towards the earlier presented performance metrics is straight-forward. This allows for real-world application of our proposed data-driven controller synthesis methods under measurement noise as has been demonstrated in~\cite{Verhoek2022_DDLPVstatefb_experiment}. 

Before we conclude this section, we want to remark on the computational complexity and conservatism of the bound on~$\epsilon$. As remarked in~\cite{dePersisTesi2020}, the bound $\frac{\alpha^2}{4+2\alpha}>\epsilon$ can be theoretically conservative, which is due to~\eqref{eq:pfth7:bound}. This means that, although there exists a stabilizing LPV controller and $\alpha$ for $\frac{\alpha^2}{4+2\alpha}>\epsilon$, the conditions of Proposition~\ref{prop:noiselmi} are not guaranteed to be feasible for this particular $\alpha$. Increasing the SNR in the data-dictionary, which can be accomplished by, e.g., increasing the magnitude of $u_k$ and $x_k$ (assuming $\varepsilon_k$ to be independent of $u_k$ and $x_k$), can make the conditions of Proposition~\ref{prop:noiselmi} feasible for larger a $\alpha$. Also note that maximizing for $\alpha$ while solving the synthesis problem, implies that the resulting controller is robust against higher noise levels. Finally, as the SDP of Proposition~\ref{prop:noiselmi} grows with $\Nd$, one can make the trade-off between robustness and computational complexity. 

{To give a comparative indication of the computational complexity that is required to solve the data-driven synthesis problems presented in this paper, we provide the dimensions of the involved LMIs in Table~\ref{tab:dims}.
\begin{table}[t]
	\centering
	\caption{{Dimensions of the LPV data-driven synthesis LMIs. Because of the symmetry of the LMIs, we only give the row/column dimension.}}
	\label{tab:dims}
	{\footnotesize \hspace*{-18pt}
	\begin{tabular}{l||c|c|c|c|c}
		  & Theorem~\ref{th:stabilizing-data-based} & Theorem~\ref{th:lqrLPV-result-data-based} & Theorem~\ref{th:H2-LPV-result-data-based} & Theorem~\ref{th:Hinf-LPV-result-data-based} & Proposition~\ref{prop:noiselmi}   \\ \hline
		Dim. of \eqref{eq:quad-matrix-inequality-multipliers:1} & $2\dnx(1+\dnp)$ & $(\dnx(1+2(1+\dnp))+\dnu)$ & $2\dnx(1+\dnp)$ & $2\dnx(2+\dnp)+\dnu$ & $(1+\dnp)(3\dnx+\Nd)$ \\
		Dim. of \eqref{eq:quad-matrix-inequality-multipliers:2} & $2\dnp\dnx$ & $2\dnp\dnx$ & $2\dnp\dnx$ & $2\dnp\dnx$ & $\dnp(3\dnx+\Nd)$ \\
		Dim. of \eqref{eq:H2-LPV-conditions-data-based} & - & - & $\dnu + \dnx(1+\dnp)$ & - & -
	\end{tabular}%
	}
\end{table}
These numbers show that, in the noise-free case, the computational complexity is mainly dependent on the size of the system. For the noisy case, however, the computational complexity also grows with $\Nd$. The synthesis LMIs in~\cite{verhoek2024decoupling, MillerSznaier2022}, where only process disturbances can be handled, do not grow with $\Nd$, which gives a computational advantage to these methods at the expense of more restrictive noise assumptions. }

\section{Simulation studies}\label{s:example}
In this section, the proposed data-driven LPV controller synthesis methods are applied in three simulation studies\footnote{For experimental results on the application of our methods on real world systems, see \cite{Verhoek2022_DDLPVstatefb_experiment}.}
to demonstrate their effectiveness and compare their achieved performance w.r.t. model-based methods. {All the examples have} been implemented using \matlab 2020a and the resulting SDPs have been solved using YALMIP~\cite{YALMIP} with the solver MOSEK~\cite{MOSEK}.
\subsection{Simulation Study 1: Data-driven scheduling independent state feedback}\label{ss:compEample}
\begin{figure}
    \centering
    \begin{subfigure}[t]{0.48\textwidth}
        \includegraphics[width=\textwidth]{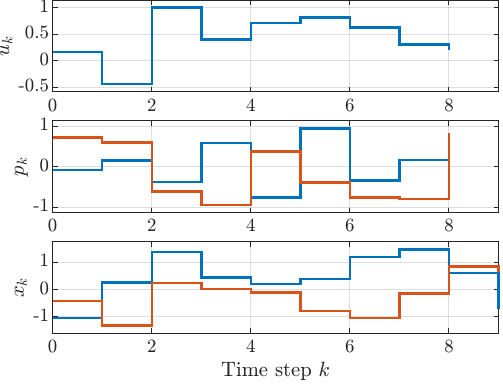}
        \caption{Data-dictionary $\dataset$ used for Simulation Study 1, consisting of 9 input-scheduling samples and 10 state samples from which $\Xf$ and $\mcG$ are constructed. For the scheduling and state plots `\legendline{mblue}' indicates the first element of the vector signals $p_k$ and $x_k$, while `\legendline{morange}' indicates the second element.}
        \label{fig:datadic_s1}
    \end{subfigure}\hfill
    \begin{subfigure}[t]{0.48\textwidth}
        \includegraphics[width=\textwidth]{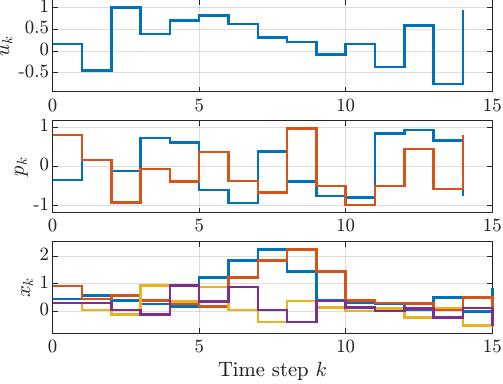}
        \caption{Data-dictionary $\dataset$ used for Simulation Study 2, consisting of 15 input-scheduling samples and 16 state samples with which $\Xf$ and $\mcG$ are constructed. For the scheduling and state plots \mbox{`\legendline{mblue}'}, \mbox{`\legendline{morange}'}, \mbox{`\legendline{myellow}'}, and \mbox{`\legendline{mpurple}'} indicate the first, second, third, and fourth element of the vector signals $p_k$ or $x_k$, respectively.}
        \label{fig:datadic_s2}
    \end{subfigure}
    \caption{Data-dictionaries used in Simulation Study 1 in Section~\ref{ss:compEample} (subfigure a) and in  Simulation Study 2 in Section~\ref{ss:stabilizingEample} (subfigure b).}
\end{figure}

In this first example, we compare our proposed data-driven methodology with model-based design in terms of stabilizing robust feedback control synthesis {and show that, our methods using purely recorded data from the system synthesize an equivalent controller with methods using the complete model of the system}. The LPV system in this example is taken from \cite{AbMaHeRo19} and can be represented as~\eqref{eq:LPVSS} with $\dnx=2$, $\dnu=1$ and $\dnp=2$,  where $A$ has affine scheduling dependence and $B$ is scheduling independent and are characterized by
\begin{align*}
	A_0 &= \begin{bmatrix}  0.2485 & -1.0355 \\ 0.8910 &   0.4065 \end{bmatrix}, & A_1 &= \begin{bmatrix} -0.0063  & -0.0938 \\ 0  &  0.0188 \end{bmatrix}, & A_2 &= \begin{bmatrix} -0.0063 & -0.0938 \\ 0  &  0.0188 \end{bmatrix}, \\  B_0 &= \begin{bmatrix} 0.3190 \\   -1.3080 \end{bmatrix}, & B_1 &= B_2 = \begin{bmatrix} 0 \\ 0 \end{bmatrix}.&&
\end{align*}
Furthermore, the scheduling set is defined as $\mathbb{P}:=[-1,\, 1]\times[-1,\, 1]$, which is polytope with four vertices. Based on the PE condition (Condition~\ref{cond:rank-cond}),  $\Nd\geq 1+(1 + \dnp)(\dnx + \dnu)=10$. Therefore, only 9 samples of input-scheduling data (the 10\tss{th} state sample is only required) are generated based on random samples from a uniform distribution: {$u_k, p_k\sim\mathcal{U}(-1,1)$}. 
By simulating the system with this $u$ and $p$ sequence and a random initial condition generated similarly, results in $\dataset$ {(depicted in Figure~\ref{fig:datadic_s1})} from which $\mcG$ is constructed. {A posteriori computation of} $\rank(\mcG)=9$ yields that Condition~\ref{cond:rank-cond} is satisfied, {i.e., $\dataset$ is PE}.

A data-driven robust (i.e., scheduling independent) quadratic-performance-optimal state-feedback controller is designed via Theorem~\ref{th:lqrLPV-result-data-based} with $Q=R=I$ (see also Remark~\ref{rem:partitioning-of-F-towards-robust}). Hence, we choose $F_{22}=0$. The resulting SDP is solved at the vertices of $\mathbb{P}$ within 0.13 seconds, and yields 
\begin{equation}\label{eq:ex1:res}
	K_0=\begin{bmatrix} 0.4832 & 0.4839 \end{bmatrix}, \  \tilde{\lyap} =\begin{bmatrix} 1.6436 & -0.4595 \\ -0.4595 & 3.0426 \end{bmatrix}, \hspace{-.5mm}
\end{equation}
where $K_0$ is obtained using~\eqref{eq:KSD-stab}. Note that choosing $F_{22}=0$ automatically renders $\bar{K}\approx 0$ up to numerical precision. %
To validate these results, the corresponding model-based conditions of Lemma~\ref{lem:LQR-LPV-model-based} have been used to design a robust state-feedback controller for the exact model of the LPV system, which resulted in a $K_0$ and $\tilde{\lyap}$ that are equivalent to~\eqref{eq:ex1:res} up to numerical precision. {Hence, this example demonstrates that, according to our theoretical results, for noise-free data and under the PE condition (Condition~\ref{cond:rank-cond}), our proposes synthesis approaches are} the data-based counterparts of model-based synthesis methods.

\subsection{Simulation Study 2: Data-driven scheduling dependent state-feedback} \label{ss:stabilizingEample}
Most of the LPV state-feedback control synthesis methodologies require \emph{constant} $B$ matrices in the LPV-SS representation of the generalized plant.
{By the authors' knowledge, the only model-based LPV controller synthesis method that can directly handle a scheduling dependent $B$ matrix is the work of~\cite{PaOl17}. The goal of this example is to show that our data-driven methods achieve the same performance as the model-based methods of~\cite{PaOl17}, despite having different underlying approaches and sources of conservatism.} We will use the benchmark LPV system used in~\cite{PaOl17} for comparison. Furthermore, we showcase the performance of the controller synthesis method using noisy data, as discussed in Section~\ref{s:noise}, and compare the results with a two-step approach (i.e., an LPV system identification step followed by a model-based controller design).

\subsubsection{LPV system and data generation}\label{sss:example:system}
The LPV system from~\cite{PaOl17} can be represented as~\eqref{eq:LPVSS} with $\dnx=4$, $\dnu=1$ and $\dnp=2$, where $A$ and $B$ have affine scheduling dependence. The matrices are characterized by
\begin{equation}\label{eq:simstudy2-system}
A_0 = \begin{bsmallmatrix} 0.8  & -0.25 &  0 &  1\\  1  & 0  & 0 &  0 \\  0 &  0 &  0.2  & 0.03\\  0  & 0 &  1  & 0 \end{bsmallmatrix},\ A_1 = \begin{bsmallmatrix} 0&0&0&0 \\ 0&0&0&0 \\ 0.8 \varrho & -0.5 \varrho &  0 &  \varrho \\ 0&0&0&0 \end{bsmallmatrix}, \ A_2 = 0_{4}, \  B_0 =\begin{bsmallmatrix} 0.5 \\ 0 \\ 0.5 \\ 0 \end{bsmallmatrix}, \ B_1 = 0_{4\times1},\  B_2 = \begin{bsmallmatrix} 0.5  \\ 0  \\ -0.5 \\  0 \end{bsmallmatrix}. 
\end{equation}
Moreover, $\mathbb{P}=[-1,\, 1]\times[-1,\, 1]$ and the parameter $\varrho \geq 0$ that is used to modify the scheduling range without affecting~$\mathbb{P}$ is chosen as in \cite{PaOl17}, i.e., $\varrho=0.53$. We generate $\dataset$ with $\Nd\geq1+(1 + \dnp)(\dnx + \dnu)=16$ by applying $u$ and $p$ sequences from $\mathcal{U}(-1,1)$ on the system (initialized with a randomly chosen initial condition $x_1$). {The resulting trajectories used for the construction of $\dataset$ are shown in Figure~\ref{fig:datadic_s2}.}
After constructing $\mcG$ from $\dataset$, the rank check in Condition~\ref{cond:rank-cond} yields  
that $\dataset$ is PE. For the example with noisy data we will discuss the data generation in Section~\ref{sss:exmp:noise}.

We will now  use the obtained $\dataset$ to synthesize a stabilizing LPV controller using Theorem~\ref{th:stabilizing-data-based} and a stabilizing LPV controller that guarantees closed-loop quadratic performance using Theorem~\ref{th:H2-LPV-result-data-based} for the considered system, which we assume to be unknown. We compare these to the model-based methods from~\cite{PaOl17}, which can also handle a scheduling dependent $B$ matrix, but require complete model-knowledge. 

\subsubsection{Stabilizing state-feedback LPV controller}\label{sss:sim2:stab}
Using Theorem~\ref{th:stabilizing-data-based}, we design a stabilizing data-driven LPV state-feedback controller based on the measured data-set. Next to the LMI constraints of Theorem~\ref{th:stabilizing-data-based}, we add $0.1<\mr{trace}(\lyap)<10$ to improve numerical conditioning of the problem. This improves the numerical stability of the inversion of $\lyap$ in~\eqref{eq:KSD-stab}. Solving the synthesis on the vertices of $\mb{P}$ yields an LPV controller of the form of~\eqref{eq:controllaw} with {its parameters computed as in~\eqref{eq:KSD-stab}}.
Additionally, we design a stabilizing LPV controller using the model-based synthesis conditions in \cite[Cor.~3]{PaOl17} using exact knowledge of the system equations and $0.1<\mr{trace}(\lyap)<10$  for numerical conditioning. The model-based design {also provides an LPV controller of the form of~\eqref{eq:controllaw}.} 

\begin{figure}
	\centering
	\begin{subfigure}[t]{0.48\linewidth}
		\includegraphics[width=\textwidth]{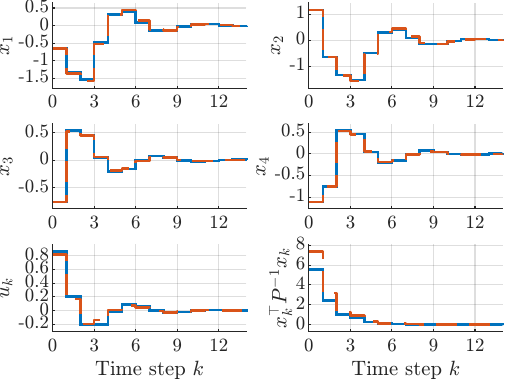}
		\caption{Simulated response of the LPV system with varying $B$ matrix in closed-loop with the stabilizing LPV controllers obtained in Section~\ref{sss:sim2:stab}. The closed-loop response with our data-driven LPV controller is represented by `\legendline{mblue}', while the closed-loop response with the model-based LPV controller obtained with the method from~\cite{PaOl17}   is depicted with `\legendlined{morange}'. We see that both controllers achieve the same performance. The differences in the results are due to the differences in the synthesis approaches.}
		\label{fig:simstudy2_1}	
	\end{subfigure}\hfill
	\begin{subfigure}[t]{0.48\linewidth}
		\includegraphics[width=\textwidth]{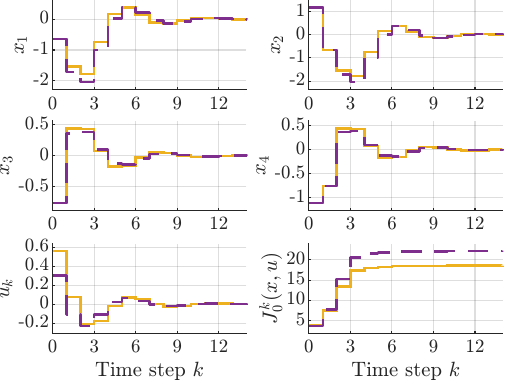}
		\caption{Simulation result of the LPV system with varying $B$ matrix in closed-loop with the LPV controllers synthesized for quadratic performance using the same cost function, see also Section~\ref{sss:sim2:perf}. The plots show that the simulation with our data-driven LPV controller (\legendline{myellow}) achieves a similar performance as the model-based LPV controller \mbox{(\legendlined{mpurple})}. Our method even achieves a lower cumulative cost as seen in the lower-right plot.}
		\label{fig:simstudy2_2}	
	\end{subfigure}
	\caption{Closed loop-responses in Simulation Study 2 with stabilizing state-feedback controllers designed in Section~\ref{sss:sim2:stab} (subfigure a), and quadratic performance optimal controllers obtained in Section~\ref{sss:sim2:perf} (subfigure b).}
\end{figure}

{We compare the performance of the two obtained controllers in closed-loop simulation, for which the results are plotted in Figure~\ref{fig:simstudy2_1}. The simulation results show that both controllers achieve the same performance (despite having different underlying approaches to solve the problem). This shows that our data-driven method is competitive with the current state-of-the-art in \emph{model-based} LPV control.}

\subsubsection{Quadratic performance state-feedback LPV controller}\label{sss:sim2:perf}
{We will perform the same comparison with LPV controllers that are synthesized for quadratic performance. We will use the extension of~\cite{PaOl17} for synthesis with quadratic performance given in~\cite[Prop.~1]{BaAbVe22} for the synthesis of the model-based LPV controller. For both designs, we choose $Q=R=I$, allowing for a quantitative comparison of the controllers.}

Next, we design a data-driven state-feedback LPV controller using the developed tools in Theorem~\ref{th:lqrLPV-result-data-based}. We solve the LMIs of Theorem~\ref{th:lqrLPV-result-data-based} as an SDP on the vertices of $\mb{P}$, where we choose the objective function as $\max \mr{trace}(\lyap)$. The solution yields an LPV controller of the form~\eqref{eq:controllaw}  with {its parameters computed as in~\eqref{eq:KSD-stab}.

For the model-based design, we use~\cite[Prop.~1]{BaAbVe22} with a constant Lyapunov matrix. Using the same objective function and tuning matrices, the model-based controller synthesis procedure is successfully solved and provided us an LPV controller of the form~\eqref{eq:controllaw} that guarantees quadratic performance. We want to emphasize that we cannot simply recast Lemma~\ref{lem:LQR-LPV-model-based} for synthesis because of the quadratic scheduling dependence in $A_\mr{CL}$ in~\eqref{eq:LQR-LPV-model-based}, which comes from the multiplication of $B(p)$ with $K(p)$. In terms of computational load, the data-based controller was synthesized within 0.65 seconds, while the model-based controller was synthesized within 0.03 seconds. This shows that the advantage of not requiring a model is traded for extra offline computation time.}

We compare the model-based LPV controller with the direct data-driven LPV controller in terms of the simulated closed-loop model response and the achieved $\htwo$-norm of the closed-loop computed on an equidistant grid\footnote{A grid-based approach is used because there is no global model-based analysis result available for LPV systems with a scheduling dependence as in~\eqref{eq:CLgenplant}. Note that, if $B$ or $K$ is scheduling independent, we can solve the model-based analysis problem in a polytopic sense using the implementations in \lpvcore.} of 250 points over $\mb{P}$. The maximum of the achieved $\htwo$-norm over the grid can be seen as an approximation of the `true' $\htwo$-norm of the LPV system. The obtained closed-loop $\htwo$ norms with the data-based and the model-based controller are $5.28$ and $5.79$, respectively, which indicates that both  controllers achieve a similar performance. {This can also be seen in the closed-loop responses that are plotted in Figure~\ref{fig:simstudy2_2}. For a quantitative comparison, we have also computed the so-called \emph{$k$-cumulative cost}:
\[ J_0^k(x,u) = \sum_{i=0}^k x_i^\top Q x_i + u_i^\top R u_i, \]
which can be seen as the finite-time approximation of the infinite-horizon cost function~\eqref{eq:InfHorcostfunction}. The $k$-cumulative cost is given in the lower-right plot of Figure~\ref{fig:simstudy2_2} and shows that the data-driven controller even achieves a lower cost than the model-based LPV controller.}

From this example, we can conclude that the proposed data-driven control synthesis machinery provides competitive controllers w.r.t. model-based approaches by using \emph{only} a few data-samples  measured from the system, rather than a full model. This considerably simplifies the overall LPV modeling and control toolchain, avoiding the two-step process of obtaining a model and then designing a model-based controller and respecting the control performance objectives in exploitation of the data. Moreover, it can cope with $p$-dependence of the $B$ matrix of the plant dynamics.

\subsubsection{Stabilizing controller synthesis with noisy measurements}\label{sss:exmp:noise}
\begin{figure}
	\centering
	\begin{subfigure}[t]{0.48\linewidth}
		\includegraphics[width=\textwidth]{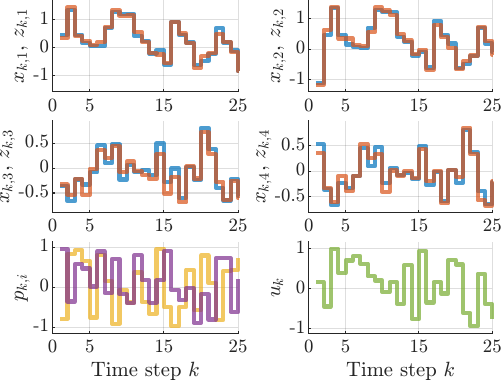}
		\caption{Noise-free ($x_{i,k}$ \legendline{mblue}) and noisy \mbox{($z_{i,k}$ \legendline{morange})} state measurements, under randomly generated scheduling $p_k$ (\legendline{myellow} and \legendline{mpurple}) and input (\legendline{mgreen}). The noisy state measurements are used in the data-dictionary~$\dataset^\varepsilon$ for Simulation Study~2.}
		\label{fig:ddic_noise}	
	\end{subfigure}\hfill
	\begin{subfigure}[t]{0.48\linewidth}
		\includegraphics[width=\textwidth]{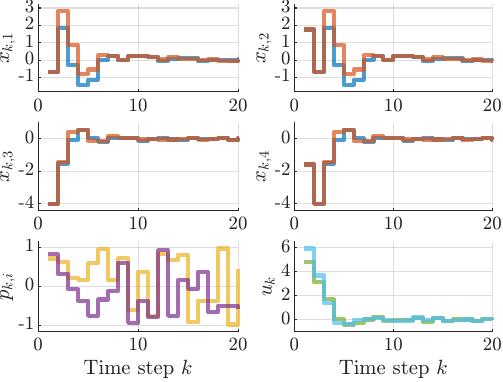}
		\caption{Closed-loop response of the LPV system under feedback law \(u_k = K(p_k) z_k\) in Simulation Study~2, where $K(p)$ is synthesized with $\dataset^\varepsilon$ using a direct approach or a two-step approach (first identification followed by model-based synthesis). The closed-loop responses in terms of the state trajectories under the same initial conditions and scheduling trajectories are plotted for the direct design~\mbox{(\legendline{mblue})} and the two-step design (\legendline{morange}), together with the corresponding scheduling (\legendline{myellow} and \legendline{mpurple}) and input signals (\legendline{mgreen} for the direct design and \legendline{mcyan} for the indirect design).}
		\label{fig:sim_noise}
	\end{subfigure}
    \caption{Closed loop-responses in Simulation Study 2 with stabilizing state-feedback controllers designed in Section~\ref{sss:exmp:noise}.}
\end{figure}

Next, we demonstrate the capabilities of the method discussed in Section~\ref{s:noise} on the considered system {and compare the results with a two-step approach}. For this part of the simulation study, we generate a new data-dictionary. We simulate the data-generating system for $\Nd=26$, while we add noise $\varepsilon_k$ to our measurements with $\varepsilon_k\sim\mc{N}(0.1,0)$, resulting in an SNR of 15 dB. The input and scheduling sequences are generated from a uniform distribution $\mc{U}(-1,1)$. The open-loop measurements used for our data-dictionary $\dataset^\varepsilon$ are shown in Figure~\ref{fig:ddic_noise}.

{Note that the noise structure is \emph{output-error} (OE), hence we can employ an LPV-OE method for the identification of the model in the two-step approach. LPV-OE identification methods are gradient-based, and with the small amount of data-points in the available $\dataset$, we have not managed to obtain a useful model from the implementation in \lpvcore~\cite{BoefCoxToth2021}. In fact, with this small number of samples, the only viable identification option is the least-squares-based LPV-ARX identification method. This approach, however, will result in a biased model estimate. This gives an advantage of the direct data-driven method over the (classic) two-step approach. Solving the LPV-ARX identification problem results in a normalized root-mean-square error of~$0.09$, i.e., a best-fit-rate of $91\%$, on a validation data set (not shown). With this model, the model-based synthesis problem is solved\footnote{In order to obtain a feasible synthesis problem, we had to relax the trace constraint on $\lyap$ from the model-based synthesis algorithm.}, resulting in an LPV controller that can stabilize the system. We will now synthesize a controller with the conditions in Proposition~\ref{prop:noiselmi} using the only data in $\dataset^\varepsilon$.}

Because of the nature of the simulation study, we can actually calculate~$\epsilon$ in Assumption~\ref{ass:noisesnr} by solving a simple minimization over~$\epsilon$ subject to $R_-R_-^\top-\epsilon\Zf\Zf^\top\negsemidef0$. This provides us with $\epsilon=0.2172$ for the obtained $\dataset^\varepsilon$. 
Next, we synthesize a stabilizing LPV controller for the system using Theorem~\ref{thm:noise}. Next to testing feasibility of~\eqref{eq:noisethmlmis}, we additionally maximize over $\alpha$ to look for the solution with the largest $\epsilon$ that still guarantees stability. {Before we give the results, we want to provide a note on the computational complexity. The SDP of the noisy synthesis problem consists of one $15\times28$ matrix equality constraint and seven LMI constraints: one for $\lyap\in\mb{S}^4$, one for $\Xi_{22}\in\mb{S}^{74}$, one for the LMI~\eqref{eq:quad-matrix-inequality-multipliers:1} of size $111\times111$ and four for the LMI~\eqref{eq:quad-matrix-inequality-multipliers:2} (because $\mb{P}$ has four vertices), each of size $74\times74$. Especially when compared to the noise-free case of Section~\ref{sss:sim2:stab} (here the LMI~\eqref{eq:quad-matrix-inequality-multipliers:1} is of size $24\times24$ and its size is independent of $\Nd$), these numbers show that the computational complexity significantly grows for synthesis with noisy data. For larger systems or big data-sets, this can be a limiting factor for achieving data-driven LPV controller design using this method.}
{In this case, however, problem is tractable and in 40 seconds} it is successfully solved on the vertices of $\mb{P}$, resulting in a stabilizing LPV controller via the parametrization~\eqref{eq:KSD-stab}. 
From the synthesis, we obtained the value $\alpha=0.0160$, which yields a $\epsilon_\mr{opt}=6.312\cdot10^{-5}$. Hence, based on the statement of Theorem~\ref{thm:noise}, the obtained controller does \emph{not} guarantee stability of the closed-loop with this noise level. However, as highlighted under Proposition~\ref{prop:noiselmi}, the bound on $\epsilon$ coming from the optimization can be theoretically conservative w.r.t. $\epsilon$. Indeed, when simulating the controller in closed-loop with the system under the feedback law \(u_k = K(p_k) z_k = K(p_k) (x_k + \varepsilon_k)\), i.e., the controller is fed with the noisy measurements, the closed-loop is stabilized and regulated to the origin without any problems, as depicted in Figure~\ref{fig:sim_noise}. In fact, if we choose our inputs for the data-generation to be larger such that the SNR goes up (i.e., $\epsilon$ shrinks), approximately the same controller is obtained for a higher $\alpha$. 

{Comparing the response of the two controller designs in Figure~\ref{fig:sim_noise}, we see that both controllers achieve a similar performance in terms of both settling time and overshoot. %
The main advantage of the two-step approach is that the synthesis procedure is not growing with the number of data-points. It is, however, highly dependent on the quality of the identified model. This is not the case for the direct data-driven design, whose computational complexity, on the other hand, grows with the number of data points. Note that the latter can be alleviated up to a certain point by selection of a short window of $\dataset$ for synthesis. In conclusion,} this example demonstrates two important aspects. First, and most important, it shows that it is possible to achieve stabilizing LPV control synthesis directly from noisy measurements without losing performance compared to the classic two-step approach. Second, it demonstrates that the currently available methods that consider noise are rather conservative and computationally expensive, motivating further research into the direction of noise handling and stochastic aspects in direct data-driven LPV control.

\subsection{Simulation Study 3: Application on the nonlinear unbalanced disc system}\label{ss:unbalanced-disc}
\begin{figure}
	\centering
	\includegraphics[height=30mm]{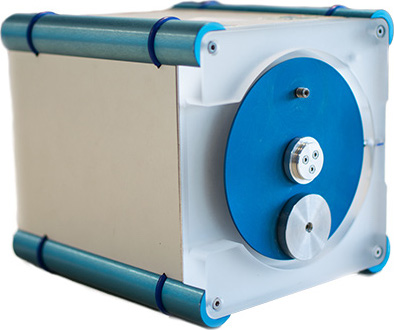}
	\caption{Unbalanced disc system, whose simulation model is used in Section~\ref{ss:unbalanced-disc}.}
	\label{fig:unbaldisc}
\end{figure}

To demonstrate  applicability of the proposed data-driven control on a real-world unstable nonlinear system, we consider the unbalanced disc setup, depicted in Figure~\ref{fig:unbaldisc}. This system consists of a DC motor that is connected to a disc containing an off-centered mass. Thus, the system behavior mimics the behavior of a rotational pendulum. {The goal of this simulation study is to show the applicability of our results for nonlinear systems and demonstrate the advantages of LPV data-driven control compared to existing LTI data-driven state-feedback control methods.}

\subsubsection{System dynamics and data-generation}
The nonlinear dynamics can be represented by the  differential equation:
\begin{equation}\label{eq:unbalanced-disc}
\ddot{\theta}(t)=-\frac{mgl}{J}\sin(\theta(t))-\frac{1}{\tau}\dot{\theta}+\frac{K_\mr{m}}{\tau}u(t),
\end{equation}
where $\theta$ is the angular position of the disc, $u$ is the input voltage to the system, which is its control input, and
 $m,g,l,J,\tau,K_\mr{m}$ are the physical parameters of the system.
\begin{table}[!ht]
	\centering
	\caption{Parameters of the unbalanced disc used in Simulation Study 3.}
	\label{tab:unbaldisc_param}
	\begin{tabular}{l||c|c|c|c|c|c|c}
		Parameter  & $g$ & $J$ & $K_\mr{m}$& $l$ & $m$  &  $\tau$  &  $T_\mr{s}$  \\ \hline
		Value & $9.8$&$2.2\cdot10^{\unaryminus 4}$ & $15.3$ &$0.42$ &$0.07$  &  $0.6$  &  $50$ \\
		Unit & $[\mr{m}\cdot\mr{s}^{\unaryminus 2}]$& $[\mr{Nm}^2]$& $[\unaryminus]$ &$[\mr{mm}]$  &  $[\mr{kg}]$  &    $[\unaryminus]$  &   $[\mr{ms}]$ 
	\end{tabular}
\end{table}
In this work, we use the physical parameters of a real setup, as given in Table~\ref{tab:unbaldisc_param}, which have been identified in \cite{KoTo19} based on measurements from the real system. It is worth mentioning that this  model was used in the past for LPV control design with successful implementation on the real hardware in \cite{AbToPeMeVeKo21}. Embedding~\eqref{eq:unbalanced-disc} as an LPV system can be established by defining the scheduling signal as $p(t) = \mr{sinc}(\theta(t))=\tfrac{\sin(\theta(t))}{\theta(t)}$ by which we can define $\mb{P}$ as $[-0.22,\; 1]$. To improve numerical conditions in the synthesis, we scaled $p$ such that $\mb{P}:=[-1,\; 1]$. 

\begin{figure}
	\centering
	\begin{subfigure}[t]{0.48\linewidth}
		\includegraphics[width=\linewidth]{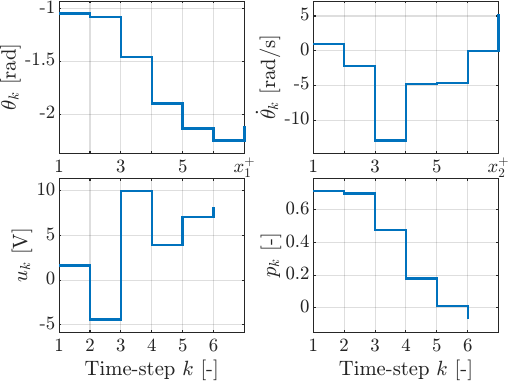}
		\caption{Data-dictionary $\dataset$ consisting of 6 input-state-scheduling samples and an additional state-advancement ($x_i^+$). The dictionary is generated by simulating the continuous-time unbalanced disc system with a random input sequence of 6 points and a random initial condition.}
		\label{fig:d}	
	\end{subfigure}\hfill
	\begin{subfigure}[t]{0.48\linewidth}
		\includegraphics[width=\linewidth]{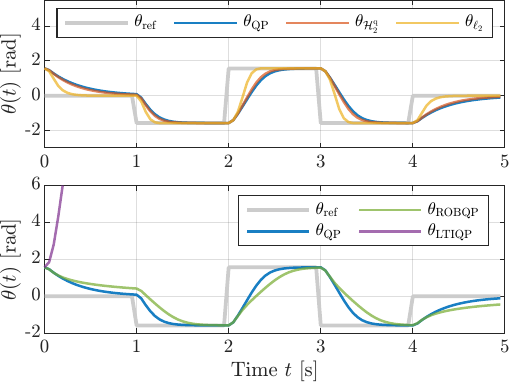}
		\caption{Simulated closed-loop responses of the unbalanced disc setup in continuous-time with the different controllers. The controllers track the reference signal (\legendline{mgray}) for the angular position $\theta$ of the disc. {The upper plot shows the difference in the performance of the designed LPV data-driven state-feedback controllers, while the lower plot provides the comparison between the  LPV design (\legendline{mblue}), robust design (\legendline{mgreen}) and the LTI design (\legendline{mpurple}).}}
		\label{fig:unbaldiscres}
	\end{subfigure}
	\caption{Used data dictionary and obtained closed loop responses of Simulation Study 3 in Section~\ref{ss:unbalanced-disc}.}
\end{figure}

For the data generation, we have implemented the nonlinear dynamical equations~\eqref{eq:unbalanced-disc} in \matlab, which are solved using an {\tt ODE45} solver at a fixed sampling time of $T_\mr{s}=0.05$. By Condition~\ref{cond:rank-cond},  we need \emph{at least} $\Nd=6$ data points. Hence, $\dataset$ is generated by exciting the system with a uniform randomly generated input of length 6, where $u$ is in the range $[-10,\,10]$. {The data is sampled in a noise-free setting from the continuous-time simulation in which the input is applied to the system via zero-order-hold.} The scheduling $p$ is a posteriori determined using the aforementioned scheduling map. The obtained data-dictionary $\dataset$ that we use to synthesize the controllers is shown in Figure~\ref{fig:d}. Note that we also depicted $\Xf$ in the upper plots of Figure~\ref{fig:d} by means of an additional state sample. For the obtained data-set, construction of $\mcG$ and computation of its rank gives $\mr{rank}(\mcG)=6$, meaning that Condition~\ref{cond:rank-cond} is satisfied.

\subsubsection{Controller synthesis}
We now use the 7 time samples in $\dataset$ that are taken from the nonlinear system to synthesize LPV state-feedback controllers for the unbalanced disc system. {For the comparison, we will synthesize five LPV state-feedback controllers:
\begin{enumerate}
    \item Using Theorem~\ref{th:lqrLPV-result-data-based}, an LPV controller with optimal quadratic performance (called the $\mr{QP}$ controller);
    \item Using Theorem~\ref{th:H2-LPV-result-data-based}, an LPV controller optimal in $\htwo$-norm performance (called the $\htwo$ controller);
    \item Using Theorem~\ref{th:Hinf-LPV-result-data-based}, an LPV controller optimal in $\ltwo$-gain performance (called the $\ltwo$ controller);
    \item Using Theorem~\ref{th:lqrLPV-result-data-based} and Remark~\ref{rem:partitioning-of-F-towards-robust}, a \emph{robust} LTI controller with optimal quadratic performance (called the $\mathrm{ROBQP}$ controller);
    \item Using \cite[Thm.~4]{dePersisTesi2020}, an LTI controller with optimal quadratic performance (denoted throughout as the $\mathrm{LTIQP}$ controller), synthesized with only input-state data.
\end{enumerate}}
Our objective is to regulate the states fast and smoothly to predefined operating points using a reasonable control input. In order to achieve this objective, {we have tuned the matrices $Q$ and $R$ for the all design problems as follows:
\begin{equation}
		Q = \begin{bmatrix} 10 & 0 \\ 0 & 2 \end{bmatrix},  \quad R = 0.1.
\end{equation}}
As in the previous examples, the cost function for the {{\rm (ROB)QP}} controllers is chosen as $\max\mr{trace}(\lyap)$, while for the $\htwo$ controller we minimize the $\htwo$-norm $\gamtwo$ during synthesis. In order to limit the aggressiveness of the $\ltwo$ controller, we modified the cost-function in the synthesis problem to $\min \gaminf+\lambda\,\mr{trace}(\lyap)$, $\lambda>0$, similar to the implementation of {\tt hinflmi} in \matlab. Note that a $\lambda\gg0$ results in larger values of $\gaminf$, i.e., the problem is regularized at the cost of performance. We choose $\lambda=0.5$. {The synthesis problems for the various controller designs are all successfully solved with the available data in $\dataset$. Next to comparison by means of simulation,} we computed the maximum $\mc{H}_2$-norm and $\mc{H}_\infty$-norm of the resulting local LTI systems of the closed-loop LPV plant over a 250-point grid over~$\mb{P}$. The results of these computations are given in Table~\ref{tab:unbaldisc_res}. 
\begin{table}
	\centering
	\caption{Obtained performance bounds provided by the synthesis methods and closed-loop analysis in Simulation Study 3.}
	\label{tab:unbaldisc_res}
	\begin{tabular}{r||c|c||c|c}
					 & $\gamma$ & $\htwo\vphantom{\Big)}$-norm & $\ltwo$-gain \\ \hline\hline
		$\mr{QP}$ controller & N/A & $11.1$ & $42.3$ \\
		$\htwo$ controller & $12.1$ & $11.2$ & $39.2$ \\
		$\ltwo$ controller  & $30.5$ & $19.5$ & $30.4$ \\
		$\mr{ROBQP}$ controller & N/A	& $17.4$ & $140$ \\
		$\mr{LTIQP}$ controller & N/A	& $\infty$ & $38.9$ \\
	\end{tabular}
\end{table}
{Note that the upper bounds $\gamma$ on the $\htwo$-norm or the $\ltwo$-gain of the closed-loop, which are provided by the synthesis procedures, are respected by the post-analysis of the unbalanced disc in closed-loop with the $\htwo$ and $\ltwo$ controllers. Moreover, it is interesting to see that both the $\mr{QP}$ controller and the $\htwo$ controller achieve a comparable closed-loop $\htwo$-norm. This is expected when comparing the definitions of the two performance notions, cf.~\eqref{eq:InfHorcostfunction} and~\eqref{eq:pf:h2:def}. %
}

\subsubsection{Simulation results}
Finally, the synthesized LPV controllers are simulated on the original nonlinear model where the control signal $u_k$ is applied to the continuous-time system in a zero-order-hold setting. Note that, by the considered state-feedback configuration, the controllers are designed for regulation to the origin. To show the capabilities of the controllers, we tested them in a setpoint control setting to achieve regulation to
\emph{different} forced equilibrium points $(x_\mr{ss},u_\mr{ss})$, as in \cite{AbToPeMeVeKo21}. In order to define the forced equilibrium points, the corresponding input steady-state values $u_\mr{ss}$ are calculated and added to the control law, which yields $u_k=K(p_k)(x_k-x_\mr{ss})+u_\mr{ss}$. We chose to switch between $x_\mr{ss} = \{ [0\ 0]^\top, \, [\tfrac{\pi}{2}\ 0]^\top, \, [-\tfrac{\pi}{2}\ 0]^\top \}$, where $x_{\mr{ss}_1}=\theta_\mr{ss}$ and $x_{\mr{ss}_2}=\dot{\theta}_{\mr{ss}}$. The simulation results for all the controllers are shown in the plots of Figure~\ref{fig:unbaldiscres}. The upper plot shows the comparison between the {\rm QP}, $\htwo$, and $\ltwo$ controllers, while the lower plot shows the comparison between the {\rm QP} controller and its LTI counterparts.

{The comparison with the different LPV controllers (upper plot in Figure~\ref{fig:unbaldiscres}) shows that the $\ltwo$ controller achieves the best tracking performance, followed by the $\htwo$ and {\rm QP} controllers. This superior tracking performance comes at the cost of more aggressive input signals, as the $\ltwo$ controller also has the most input deviations (not shown). 

The comparison of the {\rm QP} controller with the LTI controllers (lower plot in Figure~\ref{fig:unbaldiscres}) shows that the {\rm ROBQP} controller achieves robustness against all variations of $p$, i.e., the variations $\theta$, at the cost of lower performance, which is a common trade-off in robust control. The {\rm LTIQP} controller, on the other hand, does not take any variation into account and inherently assumes that the input-state data in Figure~\ref{fig:d} was coming from an LTI system. This makes that the implementation of the {\rm LTIQP} controller on the nonlinear system results in an unstable closed-loop system. This shows the advantage of LPV (data-driven) control over (robust) LTI (data-driven) control. LPV controllers are able to address directly nonlinear systems without any required trade-off between robustness and performance.}

\section{Conclusion}\label{s:conclusion}
In this work, we have derived novel direct data-driven methods that are capable of synthesizing LPV state-feedback controllers by only using information about the to-be-controlled system in the form of a persistently exciting data-set. Formulation of these results is made possible by the proposed data-driven representations of the open-loop and closed-loop behavior of the unknown LPV system. When the LPV state-feedback controllers, provided by the introduced synthesis algorithms, are connected to the unknown data-generating system, stability and performance of the closed-loop operation are guaranteed. The simulation studies show that when compared to the state-of-the-art in model-based approaches, our data-driven methods achieve the same or better performance than their model-based competitors. For the case where the data is noisy, however, the required computational load grows quadratically with $\Nd$, which can limit its applicability and thus requires further research. Additionally, we have demonstrated by means of the examples that, in line with the LPV embedding principle, the design methods can also be used for controlling nonlinear systems. We believe that our novel results can be the foundation for building a direct data-driven control framework for nonlinear systems with global stability and performance guarantees. As a future work, we aim at the extension of the synthesis methods for nonlinear systems and improving the guarantees for noisy data-sets.

\section*{Acknowledgments}
This work has received funding from the European Research Council (ERC) under the European Union's Horizon 2020 research and innovation programme (grant agreement nr. 714663), the European Union within the framework of the National Laboratory for Autonomous Systems (RRF-2.3.1-21-2022-00002), and the Deutsche Forschungs-gemeinschaft (DFG, German Research Foundation) under Project No.~419290163.

\bibliographystyle{ieeetr} 
\bibliography{references_ddlqr}

\appendix%
\section{Model-based LPV state-feedback control}\label{app:review}
In this section, the preliminaries for the proposed data-driven methods are discussed in terms of a brief summary of \emph{model-based} LPV stability and performance analysis, followed by LPV state-feedback controller synthesis.

\subsection{Model-based stability analysis}
Asymptotic stability of the LPV system~\eqref{eq:LPVSS}, i.e., boundedness and convergence of the state-trajectories to the origin under $u\equiv 0$, is guaranteed with the existence of a Lyapunov function $V: \mb{R}^{\dnx}\times\mb{P}\to\mb{R}_{\geq 0}$, where for all $p_k\in\mb{P}$,
\[ \alpha(\|x_k\|)< V(x_k,p_k) < \beta(\|x_k\|), \]
for all $x\in \mb{X}\setminus \{ 0\}$ and $V(0,p_k)=0$, where $\alpha,\beta$ are class-$K$ functions (positive, continuous, and strictly increasing functions).
We consider here Lyapunov functions that are quadratic in $x$, i.e., $V(x,p)=x^\top \lyap(p) x$  with symmetric $\lyap: \mathbb{P} \rightarrow \mb{S}^{\dnx}$ that satisfies $V(x_{k+1},p_{k+1})-V(x_k,p_k)<0$ under all $(x,p,0)\in\mathfrak{B}$. In the remainder, we will refer to $p_{k+1}$ as $p^+$ for brevity. 

Stability of the closed-loop system~\eqref{eq:closedloopsys} can be verified with the following lemma:
\begin{applemma}[Stabilizing LPV state-feedback \cite{PeWeTo18}] \label{lem:Lyapunov-LPV-model-based}
	The controller $K$ in~\eqref{eq:controllaw} asymptotically stabilizes~\eqref{eq:LPVSS}, if there exists a symmetric $\lyap:\mathbb{P}\rightarrow \mb{S}^{\dnx}$ affine in $p$ such that 
    \begin{subequations}\label{eq:Lyapunov-LPV-model-based}
		\begin{align}
			\begin{bmatrix}
				\lyap(p) & (*)^\top \\
				A_\msc{CL}(p)\lyap(p) & \lyap(p^+)
			\end{bmatrix} & \succ 0, \label{eq:Lyapunov-LPV-model-based:a} \\
			\lyap(p) & \succ 0, \label{eq:Lyapunov-LPV-model-based:b}
		\end{align}
	\end{subequations}
	for all $p, p^+\in\mathbb{P}$, with $A_\msc{CL}(p)$ as defined in~\eqref{eq:closedloopsys}.
\end{applemma}
\begin{proof}
We give it here for completeness, see also~\cite{PeWeTo18}. 
If~\eqref{eq:Lyapunov-LPV-model-based:b} holds, then $V(x,p)=x^\top \tilde\lyap(p) x > 0$ with $\lyap(p)=\tilde{\lyap}^{-1}(p)$ is true for all $(x,p)\in\mathfrak{B}_\msc{CL}$, where 
\[ \mathfrak{B}_\msc{CL} = \{ (x,p) \mid (x,p,K(p)x)\in\mathfrak{B}\}. \]
Furthermore,~\eqref{eq:Lyapunov-LPV-model-based:a} under~\eqref{eq:Lyapunov-LPV-model-based:b} is equivalent with
\[
A_\msc{CL}^\top(p)\tilde{\lyap}(p^+)A_\msc{CL}(p)  -  \tilde{\lyap}(p)  \negdef 0,\quad\forall    p, p^+\in\mathbb{P}, 
 \]
 in terms  of the Schur complement. This implies that $V(x^+,p^+)-V(x,p) < 0$ for all vectors $p, p^+\in\mathbb{P}$ and $x\in\R^{n_\mr{x}}\setminus\{0\}$ and hence under all $(x,p)\in\mathfrak{B}_\msc{CL}$. Based on the fact that asymptotic stability of~\eqref{eq:closedloopsys} is equivalent with the existence of a quadratic Lyapunov function fulfilling the above conditions \cite{PeWeTo18}, i.e.,~\eqref{eq:closedloopsys} is stable if~\eqref{eq:Lyapunov-LPV-model-based} hold for all $p, p^+\in\mathbb{P}$, concluding the proof.
\end{proof}

 \subsection{Quadratic performance analysis}
 
Beyond stability, design of a controller~\eqref{eq:controllaw} can be used to ensure performance specifications on the closed-loop behavior. Performance of~\eqref{eq:closedloopsys} can be expressed in various forms, such as the quadratic infinite-time horizon cost in~\eqref{eq:InfHorcostfunction}.
The following condition can be used to test whether a given $K$ stabilizes~\eqref{eq:LPVSS} and achieves the smallest possible bound on~\eqref{eq:InfHorcostfunction} under all $p\in\mb{P}^{\Z}$, i.e., achieves a minimal $\sup_{(x,p)\in\mathfrak{B}_\msc{CL}} J(x,K(p)x)$.

\begin{applemma}[Optimal quadratic performance~LPV state-feedback~\cite{rotondo2015linear}] \label{lem:LQR-LPV-model-based} 
	The controller $K$ in~\eqref{eq:controllaw} asymptotically stabilizes~\eqref{eq:LPVSS} and achieves the minimum of $\sup_{(x,p)\in\mathfrak{B}_\msc{CL}} J(x,K(p)x)$, if there exists a symmetric $\lyap:\mb{P}\to\mb{S}^{\dnx}$ such that $\sup_{p\in\mb{P}}\mr{trace}(\lyap(p))$  is minimal among all possible choices of $\lyap,K$ {that satisfy}
    \begin{subequations}\label{eq:LQR-LPV-model-based}
		\begin{align}
			\begin{bmatrix}
			\lyap(p) & (*)^\top & (*)^\top  & (*)^\top \\
			A_\msc{CL}(p)\lyap(p) & \lyap(p^+) & 0  & 0 \\
			Q^{\frac{1}{2}}\lyap(p) & 0 & I_{n_\mr{x}} & 0 \\
			R^{\frac{1}{2}}K(p)\lyap(p) & 0 & 0 & I_{n_\mr{u}} 
			\end{bmatrix}&\succ 0, \label{eq:LQR-LPV-model-based:a}
			\\
			\lyap(p) &\succ 0,
		\end{align}
	\end{subequations}
	for all $p, p^+\in\mathbb{P}$. 
\end{applemma}
\begin{proof}
    If there exists a quadratic Lyapunov function $V(k):= x_k^\top \tilde\lyap(p_k)x_k>0$, with $\tilde\lyap(p_k) \posdef 0$ for all $p_k\in\mb{P}$, such that
    \begin{equation}\label{eq:proof:2:dev}
        V(k+1) - V(k) \le -(x_k^\top Q x_k + u_k^\top R u_k), \quad \forall p_k\in\mb{P} 
    \end{equation}
    with $u_k=K(p_k)x_k$ then, as the right-hand side is negative semidefinite due to $Q\succeq 0,R\succ 0$,
    $K$ is an asymptotically stabilizing controller for the LPV system given by~\eqref{eq:LPVSS}, ensuring that $x_k\to0$ as $k\to\infty$.
    Rewriting~\eqref{eq:proof:2:dev}, we get
    \begin{equation}\label{eq:pf:lqr1}
        x_{k+1}^\top\tilde\lyap(p_{k+1})x_{k+1} - x_{k}^\top\tilde\lyap(p_{k})x_{k}  \le - x_k^\top\left( Q + K^{\top}(p_k)R K(p_k)\right)x_k.
    \end{equation}
    Summing all terms from 0 to $\infty$ yields
    \begin{equation}\label{eq:pf:telescope}
        \sum_{k=0}^\infty x_{k+1}^\top\tilde\lyap(p_{k+1})x_{k+1} - x_{k}^\top\tilde\lyap(p_{k})x_{k} \le -J(x,K(p)x).
    \end{equation}
    If~\eqref{eq:pf:lqr1} holds, then, in terms of~\eqref{eq:proof:2:dev}, $x_k\to0$ as $k\to\infty$, and the telescopic sum on
    the left-hand side of~\eqref{eq:pf:telescope} reduces to
    $-x_0^\top\tilde\lyap(p_0)x_0$, i.e.,
    $ x_0^\top\tilde\lyap(p_0)x_0 \ge J(x,K(p)x)$.
    Hence, $x_0^\top\tilde\lyap(p_0)x_0$ is an upper bound on $J$, given that~\eqref{eq:pf:lqr1} holds. {Inequality~\eqref{eq:pf:lqr1} is implied by} %
    \begin{equation}\label{eq:pf:lqr2}
    A_\msc{CL}^\top(p)\tilde{\lyap}(p^+)A_\msc{CL}(p)  -  \tilde{\lyap}(p) +Q+K^{\hspace{-0.2mm}\top}\hspace{-0.9mm}(p) R K(p) \negsemidef 0,
    \end{equation}
   for all $p,p^+\in\mb{P}$.
    Then, %
    minimizing $J$ over all possible state and scheduling trajectories, i.e.,  $\sup_{(x,p)\in\mathfrak{B}_\msc{CL}} J(x,K(p)x)$, can be rewritten as
    \begin{subequations}\label{eq:pf:lqr:opt}
    \begin{align}
        \min_{\tilde\lyap,K} & \quad   x_0^\top\tilde\lyap(p_0)x_0 \\
        \text{s.t.} & \quad \forall p,p^+\in\mb{P}: \tilde\lyap(p)\posdef0 \text{ and~\eqref{eq:pf:lqr2} holds}, \label{eq:pf:lqr:condition}
    \end{align}
    \end{subequations}
    for all possible initial conditions $x_0\in\mathbb{R}^{n_\mr{x}}$ and $p_0\in\mathbb{P}$. By using the eigendecomposition of $\tilde\lyap(p)$,~\eqref{eq:pf:lqr:opt} 
    is equivalent with minimizing $\mr{trace}(\tilde\lyap(p))$ subject to~\eqref{eq:pf:lqr:condition} over all $p\in\mb{P}$. Moreover, applying the Schur complement on~\eqref{eq:pf:lqr2} w.r.t. $\mr{blkdiag}(\tilde{\lyap}(p^+),I_{n_\mr{x}},I_{n_\mr{u}})$, followed by a congruence transformation $\mr{blkdiag}(\lyap(p),I_{n_\mr{x}},I_{n_\mr{x}},I_{n_\mr{u}})$ on the resulting matrix inequality, where $\lyap(p)=\tilde{\lyap}^{-1}(p)$,  yields
    \begin{align*}
        \min_{\lyap,K} &  \quad\sup_{p\in\mb{P}}\ \mr{trace}(\lyap(p)) \\
        \text{s.t.} & \quad \forall p,p^+\in\mb{P}: \text{~\eqref{eq:LQR-LPV-model-based} holds,} 
    \end{align*}
    completing the proof.
\end{proof}

\subsection{$\htwo$ performance analysis}\label{app:h2}
Using the definition of the $\htwo$-norm in~\eqref{eq:h2normdef}, the following lemma, which is a slight modification of~\cite[Lem.~1]{deCaigny2010_H2LPV}, allows to find a bound $\gamma$ on the $\htwo$-norm of the closed-loop system and guarantee stability of~\eqref{eq:CLgenplant}.

\begin{applemma}[$\htwo$-norm performance LPV state-feedback]\label{lem:H2-LPV-conditions-model-based} {The controller $K$ in~\eqref{eq:controllaw} asymptotically stabilizes~\eqref{eq:LPVSS} and achieves 
	an $\htwo$-norm of~\eqref{eq:CLgenplant} that is} less than {a given} $\gamtwo>0$, if there exist symmetric matrices $\lyap:\mathbb{P}\rightarrow\mb{S}^{\dnx}$ and $S\in\mb{S}^{\dnu}$ such that
	\begin{subequations}\label{eq:H2-LPV-conditions-model-based}
	\begin{align}
    	\begin{bmatrix}
    	\lyap(p^+)-I_{n_\mr{x}}  & A_\msc{CL}(p)\lyap(p)\\
    	(\ast)^\top & \lyap(p)
    	\end{bmatrix} &\succ 0, \label{eq:H2-LPV-conditions-model-based:a} \\
    	\begin{bmatrix}
    	S & R^{\frac{1}{2}}K(p)\lyap(p) \\
    	(\ast)^\top & \lyap(p)
    	\end{bmatrix} &\succ 0, \label{eq:H2-LPV-conditions-model-based:b} \\
    	\mr{trace}(Q{\lyap}(p))+\mr{trace}(S) &\prec \gamtwo^2, \label{eq:H2-LPV-conditions-model-based:c}
	\end{align}
	\end{subequations}
	for all $p,p^+\in\mb{P}$.
\end{applemma}
\begin{proof}
Applying the result of \cite[Lem.~1]{deCaigny2010_H2LPV} to~\eqref{eq:CLgenplant}, denoted by $\Sigma_{\msc{CL}}$, the $\htwo$-norm can be written as
\begin{equation}\label{eq:pf:h2:def}
	\hspace{-1mm}\|\Sigma_{\msc{CL}}\|_{\htwo}^2= \limsup_{N\to\infty} \ \tfrac{1}{N}\sum_{k=0}^N \mr{trace}\left(C(p_k)\bar\lyap(p_k)C^{\hspace{-0.2mm}\top}\hspace{-0.9mm}(p_k)\right),
\end{equation}
where $\bar\lyap(p_k)$ is the controllability Gramian that satisfies
\begin{equation} \label{eq:proof:3:dev:1}
	A_\msc{CL}(p_k)\bar{\lyap}(p_k)A_\msc{CL}^\top(p_k) - \bar{\lyap}(p_{k+1}) + I_{n_\mr{x}} = 0
\end{equation}
with $\bar{\lyap}(p_0)=0$. Inequality~\eqref{eq:H2-LPV-conditions-model-based:a} implies that ${\lyap}(p) \posdef 0$ and
\begin{equation}
	A_\msc{CL}(p){\lyap}(p)A_\msc{CL}^\top(p) - {\lyap}(p^+) + I_{n_\mr{x}} \negdef 0,
\end{equation}
which ensures asymptotic stability of~\eqref{eq:CLgenplant}. Furthermore, there exists a $W(p)=W^\top(p)\posdef 0$ such that
\begin{equation}
	A_\msc{CL}(p){\lyap}(p)A_\msc{CL}^\top(p) - {\lyap}(p^+) + I_{n_\mr{x}} + W(p) = 0,
\end{equation}
which implies~\eqref{eq:proof:3:dev:1} for all $k\ge0$ with $\lyap\posdef\bar\lyap$. Moreover,~\eqref{eq:H2-LPV-conditions-model-based:b} implies that
\begin{equation}
	S \posdef R^{\frac{1}{2}}K(p)\lyap(p)K^{\hspace{-0.2mm}\top}\hspace{-0.9mm}(p)R^{\frac{\top}{2}}.
\end{equation}
Substituting the latter in~\eqref{eq:H2-LPV-conditions-model-based:c} gives
\begin{equation}
	\mr{trace}(Q\lyap(p)) + \mr{trace}(S)>
	\mr{trace}(Q\lyap(p)) + \mr{trace}(R^{\frac{1}{2}}K(p)\lyap(p)K^{\hspace{-0.2mm}\top}\hspace{-0.9mm}(p)R^{\frac{\top}{2}}),
\end{equation}
where the right-hand side can be rewritten using the cyclic property of the trace as
\begin{equation}
	\mr{trace}(C(p)\lyap(p)C^{\hspace{-0.2mm}\top}\hspace{-0.9mm}(p)),
\end{equation}
with $C(p)=\begin{bmatrix} (Q^{\frac{1}{2}})^\top & (R^{\frac{1}{2}}K(p_k))^\top\end{bmatrix}^\top$. %
Hence, finding a $\gamma$ for which~\eqref{eq:H2-LPV-conditions-model-based} holds for all $p\in\mb{P}$ ensures via~\eqref{eq:H2-LPV-conditions-model-based:c} that~\eqref{eq:pf:h2:def} is  upper bounded by $\gamma$, concluding the proof.
\end{proof}

\subsection{$\ltwo$-gain performance analysis}
The following result allows to analyze stability and $\ltwo$-gain performance, as introduced in Section~\ref{ss:ltwosyn}, of the closed-loop system~\eqref{eq:CLgenplant}.

\begin{applemma}[$\ltwo$-gain performance LPV state-feedback]\label{lem:Hinf-LPV-conditions-model-based} 
	The controller $K$ in~\eqref{eq:controllaw} asymptotically stabilizes~\eqref{eq:LPVSS} and achieves an $\ltwo$-gain of~\eqref{eq:CLgenplant} that is less than a given $\gaminf >0$, if there exists a symmetric matrix $\lyap:\mathbb{P}\rightarrow\mb{S}^{\dnx}$ such that
	\begin{subequations}\label{eq:Hinf-LPV-conditions-model-based}
		\begin{align}
			\begin{bmatrix}
			\lyap(p) & (*)^\top      & (*)^\top & (*)^\top & 0 \\
			A_\msc{CL}(p){\lyap}(p)  & \lyap(p^+)  & 0 & 0 & I_{\dnx}\\
			Q^{\frac{1}{2}}\lyap(p) & 0  & \gaminf I_{n_{\mr{x}}}  & 0 & 0 \\ 
			R^{\frac{1}{2}}K(p)\lyap(p) & 0  & 0 & \gaminf I_{n_{\mr{u}}}  & 0  \\
			0  &  I_{n_{\mr{x}}}   &  0  & 0  & \gaminf I_{n_{\mr{x}}}
			\end{bmatrix} &\succ 0, \label{eq:Hinf-LPV-conditions-model-based:a}\\
			 {\lyap}(p)  & \succ  0,\label{eq:Hinf-LPV-conditions-model-based:b}
		\end{align}
	\end{subequations}
	for all $p,p^+\in\mb{P}$.
\end{applemma}
\begin{proof}
	Following standard formulation, e.g., \cite{GaAp94}, for $\ltwo$-gain performance for the closed-loop LPV system~\eqref{eq:CLgenplant}, the following condition should hold for all $p,p^+\in\mb{P}$:
	\begin{equation}
	\begin{bmatrix}
	(*)^\top\tilde{\lyap}(p^+)A_\msc{CL}(p)-\tilde{\lyap}(p)
	& (*)^\top\\
	\tilde{\lyap}(p^+)A_\msc{CL}(p) & \tilde{\lyap}(p^+)-\gaminf I_{n_{\mr{x}}}
	\end{bmatrix}
	+\frac{1}{\gaminf}\begin{bmatrix}*\end{bmatrix}^\top\begin{bmatrix}
	Q^{\frac{1}{2}} &0 \\ R^{\frac{1}{2}}K(p) & 0
	\end{bmatrix} 
	\prec 0
	\end{equation}
	with $\tilde{\lyap}^\top(p)=\tilde{\lyap}(p)\succ 0$. The above matrix inequality   can be rewritten as 
	\begin{equation}
	0 \prec \begin{bmatrix}
	\tilde{\lyap}(p) & 0\\ 0 & \gaminf I_{n_{\mr{x}}}
	\end{bmatrix} - 
	\begin{bmatrix}
	*
	\end{bmatrix}^\top
	\begin{bmatrix}
	\tilde{\lyap}^{-1}(p^+) &0 &0 \\
	0 & \tfrac{1}{\gaminf}I_{n_\mr{x}} &0 \\
	0 & 0 & \tfrac{1}{\gaminf}I_{n_\mr{u}}
	\end{bmatrix}
	\begin{bmatrix}
	\tilde{\lyap}(p^+)A_\msc{CL}(p) & \tilde{\lyap}(p^+)\\
	Q^{\frac{1}{2}}  & 0  \\ R^{\frac{1}{2}}K(p) & 0 
	\end{bmatrix}.
	\end{equation}  
	Now, applying the Schur complement with respect to $\mr{blkdiag}(\tilde{\lyap}^{-1}(p^+),\tfrac{1}{\gaminf}I_{n_\mr{x}},\tfrac{1}{\gaminf}I_{n_\mr{u}})$ and the congruence
	transformation $\mr{blkdiag}({\lyap}(p),I_{n_{\mr{x}}},
	{\lyap}(p^+),I_{n_{\mr{x}}},I_{n_{\mr{u}}})$ on the resulting matrix inequality, where $\lyap(p)=\tilde{\lyap}^{-1}(p)$, yields the condition
	\begin{equation*}
		\begin{bmatrix}
		\lyap(p) & 0    & (*)^\top  & (*)^\top & (*)^\top \\
		0 & \gaminf I_{n_{\mr{x}}} & I_{\dnx}   & 0 & 0  \\
		A_\msc{CL}(p){\lyap}(p) & I_{n_{\mr{x}}} & \lyap(p^+) & 0 & 0 \\
		Q^{\frac{1}{2}}\lyap(p) & 0 & 0 & \gaminf I_{n_{\mr{x}}} & 0\\ 
		R^{\frac{1}{2}}K(p)\lyap(p) & 0 & 0 & 0 & \gaminf I_{n_{\mr{u}}}
		\end{bmatrix} \succ 0. 	
	\end{equation*}
	Finally, pre- and post-multiplication of the above matrix inequality by the permutation matrix 
	\[
	P=\begin{bmatrix}
	I_{n_{\mr{x}}} & 0 & 0 &  0 & 0\\
	0 & 0 & 0 &  0 &I_{n_{\mr{x}}} \\
	0 & I_{n_{\mr{x}}} & 0 &  0 &0\\
	0 & 0 & I_{n_{\mr{x}}} &  0 &0\\
	0 & 0 &   0  & I_{n_{\mr{u}}} &0\\
	\end{bmatrix}
	\]
	and its transpose, respectively, yields condition~\eqref{eq:Hinf-LPV-conditions-model-based:a}.
\end{proof}

\subsection{Model-based controller synthesis}\label{ss:analysis2synthesis}
When $K$ is \emph{not} known, the matrix inequality conditions in Lemmas~\ref{lem:Lyapunov-LPV-model-based}--\ref{lem:Hinf-LPV-conditions-model-based} are \emph{nonlinear} in the decision variables (controller parameters and $P$ characterizing the Lyapunov/storage function) and provide an infinite number of inequalities that need to be satisfied for every point in $\mb{P}$. 
To resolve these problems and recast these conditions to tractable controller synthesis methods, the change of variables $Y(p)=K(p)\lyap(p)$ can be applied in the conditions~\eqref{eq:Lyapunov-LPV-model-based:a},~\eqref{eq:LQR-LPV-model-based:a},~\eqref{eq:H2-LPV-conditions-model-based:a},~\eqref{eq:H2-LPV-conditions-model-based:b} and~\eqref{eq:Hinf-LPV-conditions-model-based:a}. Furthermore, the dependence on~$p$ must be defined for $K$ and ${\lyap}$ (i.e., for $Y(p)$), which has an important impact on the complexity of the synthesis problem and the achievable control performance. A natural choice for the scheduling dependence of $K$ is to assume static-affine dependence on $p$ as in~\eqref{eq:controllaw:dependencyK}.
On the other hand, the choice of the scheduling dependence of ${\lyap}$ is not trivial, as discussed in \cite{ApAd98}, and it is often accomplished in terms of the choice of $Y$. These considerations allow to convert the corresponding LPV state-feedback controller synthesis problems into the minimization of a linear cost, subject to constraints defined by an infinite set of LMIs.
The last step in making the synthesis problems tractable is reducing the set of constraints to a finite number of LMIs, such that the resulting problem can be efficiently solved as an SDP using off-the-shelf solvers. There are multiple methods available to accomplish this. As can be seen in the main body of this paper, we use the full-block $\mc{S}$-procedure, see Lemma~\ref{lem:full-block-S-procedure}, to recast the synthesis problems to an SDP of finite size.

\end{document}